\documentstyle[12pt]{article}
\def\journal{\topmargin .3in	\oddsidemargin .5in
	\headheight 0pt	\headsep 0pt
	\textwidth 5.625in 
	\textheight 8.25in 
	\marginparwidth 1.5in
	\parindent 2em
	\parskip .5ex plus .1ex		\jot = 1.5ex}
\journal

\catcode`\@=11
\def\marginnote#1{}
\catcode`\@=11
\def\section{\@startsection {section}{1}{0pt}{-3.5ex plus -1ex minus
 -.2ex}{2.3ex plus .2ex}{\raggedright\large\bf}}
\catcode`\@=12
\newskip\humongous \humongous=0pt plus 1000pt minus 1000pt
\def\caja{\mathsurround=0pt}
\def\eqalign#1{\,\vcenter{\openup1\jot \caja
	\ialign{\strut \hfil$\displaystyle{##}$&$
	\displaystyle{{}##}$\hfil\crcr#1\crcr}}\,}
\newif\ifdtup
\def\panorama{\global\dtuptrue \openup1\jot \caja
	\everycr{\noalign{\ifdtup \global\dtupfalse
	\vskip-\lineskiplimit \vskip\normallineskiplimit
	\else \penalty\interdisplaylinepenalty \fi}}}
\def\eqalignno#1{\panorama \tabskip=\humongous
	\halign to\displaywidth{\hfil$\displaystyle{##}$
	\tabskip=0pt&$\displaystyle{{}##}$\hfil
	\tabskip=\humongous&\llap{$##$}\tabskip=0pt
	\crcr#1\crcr}}
\def\N{{\rm I\!N}}
\font\fivesans=cmss10 at 4.61pt
\font\sevensans=cmss10 at 6.81pt
\font\tensans=cmss10
\newfam\sansfam
\textfont\sansfam=\tensans\scriptfont\sansfam=\sevensans\scriptscriptfont
\sansfam=\fivesans
\def\sans{\fam\sansfam\tensans}
\def\Z{{\mathchoice
{\hbox{$\sans\textstyle Z\kern-0.4em Z$}}
{\hbox{$\sans\textstyle Z\kern-0.4em Z$}}
{\hbox{$\sans\scriptstyle Z\kern-0.3em Z$}}
{\hbox{$\sans\scriptscriptstyle Z\kern-0.2em Z$}}}}
\def\a{\alpha}
\def\b{\beta}
\def\g{\gamma}
\def\d{\delta}

\def\m{\eta}

\def\p{\pi}
\def\r{\rho}
\def\s{\sigma}
\def\hh{\tilde h}
\def\l{\lambda}

\def\ra{\rightarrow}
\def\bz{\bar{z}}
\def\bc{\bar{\g}}
\def\bu{\bar{u}}
\def\bw{\bar{w}}
\def\bi{\bar{\imath}}
\def\bj{\bar{\jmath}}

\def\bR{\bar{R}}
\def\tL{\tilde{L}}
\def\tT{\tilde{T}}
\def\tD{\tilde{\Delta}}
\def\pa{\partial}
\def\bpa{\bar{\partial}}
\def\f{\phi}
\def\tf{\bar{\phi}}
\def\bA{\bar{A}}

\def\ve{\vert}
\def\J{{\cal J}}
\def\df{\delta \phi}
\def\tdf{\tilde{\delta}\phi}

\def\bY{\bar{Y}}
\def\bq{\bar{q}}
\def\bJ{\bar{{\cal J}}}

\def\lra{\leftrightarrow}
\def\H{{\cal H}}
\def\C{{\cal C}}
\def\G{{\cal G}}
\def\bg{{\rm {\bf g}}}
\def\lg{\l_{\bg_x}}
\def\F{{\cal F}}

\def\ha{\hat{\a}}
\def\hb{\hat{\b}}
\def\ba{{\rm {\bf a}}}
\def\bb{{\rm {\bf b}}}
\def\Ga{\Gamma}
\def\Fi{\Phi}
\def\xx{\hbox{ }^*_*}

\def\D{\Delta}
\def\T{{\cal T}}
\def\C{{\cal C}}

\def\ref#1{$^{#1)}$}
\begin{document}
\begin{titlepage}
\begin{center}
July 1992           \hfill    LBL-32619 \\
                    \hfill    UCB-PTH-92/24 \\
                    \hfill    BONN-HE-92/21
\vskip .1in

{\large \bf Ward Identities for Affine-Virasoro Correlators}
\footnote{The work of MBH was supported in part by the Director, Office of
Energy Research, Office of High Energy and Nuclear Physics, Division of
High Energy Physics of the U.S. Department of Energy under Contract
DE-AC03-76SF00098 and in part by the National Science Foundation under
grant PHY90-21139. The work of NO was supported in part by the Deutsche
Forschungsgemeinschaft.}

\vskip .2in
M.B. Halpern
\footnote{e-mail: HALPERN\%THEORM.HEPNET@LBL.GOV, THEORY::HALPERN}
\vskip .05in
{\em  Department of Physics, University of California\\
      Theoretical Physics Group, Lawrence Berkeley Laboratory\\
      Berkeley, California 94720 \\
       USA}
\vskip .1in
N.A. Obers
\footnote{e-mail: OBERS@PIB1.PHYSIK.UNI-BONN.DE, 13581::OBERS}
\vskip .05in
{\em Physikalisches Institut der Universit\"at Bonn \\
Nu{\ss}allee 12, D-5300 Bonn 1 \\
Germany}
\end{center}

\vskip .05in

\begin{abstract}
Generalizing the Knizhnik-Zamolodchikov equations, we derive a hierarchy of
non-linear Ward identities for affine-Virasoro correlators. The hierarchy
follows from null states of the Knizhnik-Zamolodchikov type and the assumption
of factorization, whose consistency we verify at an abstract level. Solution of
the equations requires concrete factorization ans\"atze, which may vary over
affine-Virasoro space. As a first example, we solve the non-linear equations
for the coset constructions, using a matrix factorization. The resulting coset
correlators satisfy first-order linear partial differential equations whose
solutions are the coset blocks defined by Douglas.
\end{abstract}

\end{titlepage}

\newpage
\renewcommand{\thepage}{\arabic{page}}
\setcounter{page}{1}
\setcounter{footnote}{1}
\section{Introduction}

Affine-Virasoro constructions are Virasoro operators constructed with the
currents $J_a,\,a=1,\ldots,{\rm dim}\,g$ of affine $g$. Here is a brief
history of these constructions.

Affine Lie algebra, or current algebra on $S^{1}$, was discovered independently
in mathematics \cite{km} and physics \cite{bh}. The first concrete
representation \cite{bh} was untwisted $SU(3)_1$, obtained with world-sheet
fermions \cite{bh,r} in the construction of current-algebraic spin and internal
symmetry on the string \cite{bh}. Examples of affine-Sugawara\footnote{The
Sugawara-Sommerfeld model \cite{ss} was in four dimensions on the algebra of
fields. The first affine-Sugawara constructions, on affine Lie algebra, were
given by Bardak\c ci and Halpern \cite{bh,h1} in 1971.} constructions
\cite{bh,h1} and coset constructions \cite{bh,h1} were also given in the first
string era, as well as the vertex operator construction of fermions and
untwisted $SU(n)_1$ from compactified spatial dimensions \cite{h2,bhn}. The
generalization of these constructions \cite{lw,wit1,kz,go} and their
application to the heterotic string \cite{gh} mark the beginning of the present
era. See Refs. \cite{fh,h3} for further historical remarks on these early
affine-Virasoro constructions.

The general affine-Virasoro construction \cite{hk,rus}
$$ T(L) = L^{ab}\xx J_a J_b \xx \eqno(1.1) $$
is summarized by the Virasoro master equation \cite{hk,rus} for the inverse
inertia tensor $L^{ab}=L^{ba}$. A generalized master equation including $\pa J$
terms \cite{hk} has also been obtained, as well as the superconformal master
equation \cite{sme}, which collects the superconformal solutions of the
Virasoro master equation.

Here is an overview of the solution space, called affine-Virasoro space, of
the Virasoro master equation.
\begin{enumerate}
\item[a)] The standard rational conformal field theories are contained
in the affine-Sugawara nests \cite{nuc}, which include the
affine-Sugawara constructions \cite{bh,h1,wit1,kz}, the coset constructions
\cite{bh,h1,go} and the nested coset  constructions \cite{wit2,nuc}.
\item[b)] The master equation has a very large number of solutions, e.g.
approximately 1/4 billion on each level of affine $SU(3)$, and exponentially
larger numbers on larger manifolds \cite{nuc}. Most of these constructions are
new, and many new solutions have been found in closed form, including large
numbers of unitary solutions with irrational central charge
\cite{nuc,slg,st,hl,gt,mb,sin,ggt,lh}. As examples, the value at level
5 of $SU(3)$ \cite{hl}
$$ c\left( (SU(3)_5)^{\#}_{D(1)}\right) =2 \left(1-{1 \over \sqrt{61}}\right)
\simeq 1.7439 \eqno(1.2)$$
is the lowest unitary irrational central charge yet observed, while the
simplest
exact unitary irrational level-families yet obtained are the
$rs$-superconformal set with central charge \cite{sin}
$$ c(SU(n)^{\#}_x[m(N=1),rs])={6nx \over nx + 8\sin^2(rs \pi/n) } \eqno(1.3)$$
where $r,s \in \N$ and $x$ is the level of affine $SU(n)$. Ref. \cite{lh} gives
the most recent list of exact unitary solutions with irrational central charge.
Large classes of unitary irrational solutions have also been studied by
high-level expansion \cite{hl}, which remains the most powerful tool so far
developed for the study of the space.
\item[c)] The generic conformal field theory has irrational central charge, and
rational central charge is rare in the space of unitary conformal field
theories. Indeed, the standard rational conformal field theories live in the
much larger space of Lie $h$-invariant conformal field theories \cite{lh},
which are themselves quite rare. Large numbers of candidates for new rational
conformal field theories, beyond the affine-Sugawara nests, have also been
found \cite{ks,gep,nsc}.
\item[d)] Partial classification of affine-Virasoro space has been achieved
with
graph theory and generalized graph theories
\cite{gt,mb,sme,nsc,ssc,sin,ggt,lh}.
Moreover, the master equation generates the graph theories on the group
manifolds in such a way that unsuspected Lie group structure, called
Generalized Graph Theory on Lie $g$ \cite{mb,ggt}, is seen in each of the
graph theories. The interested reader should consult Ref. \cite{ggt}, which
axiomatizes the subject.
\item[e)] Large as they are, the graph theories so far cover only very small
regions of affine-Virasoro space. Enough has been learned, however, to see that
all known exact solutions are special cases with relatively high symmetry,
whereas the generic solution is completely asymmetric \cite{gt}. In this
circumstance, an exact general solution of the master equation seems beyond
hope.
\end{enumerate}

We also mention a number of other developments in the program, including
geometric identification \cite{gme} of the master equation as an Einstein-like
system on the group manifold, a world-sheet action \cite{gva} for the generic
affine-Virasoro construction on simple $g$, and the exact C-function \cite{cf}
and C-theorem on affine-Virasoro space. See also Ref. \cite{rd}, which gives an
introductory review of developments in the Virasoro master equation.

It is clear that the Virasoro master equation is the first step in the study
of irrational conformal field theory, but how are we to obtain the correlators
of such theories$\,?\,$Most of the computational methods of conformal field
theory \cite{bpz,df} are based on chiral algebras \cite{rns,w} and their
corresponding chiral null states, a situation of relatively high symmetry which
cannot be generic in affine-Virasoro space. On the other hand, each
affine-Virasoro construction has null states of the Knizhnik-Zamolodchikov
(KZ) type \cite{kz}, if only we can learn to exploit them.

The context for this development was given in the 1989 paper ``Direct Approach
to Operator Conformal Constructions'' \cite{h3}, by one of the present authors.
The central point is that affine-Virasoro constructions come in commuting
K-conjugate pairs \cite{bh,h1,go,hk}, which naturally form {\em biconformal
field theories}. In these systems, the natural
analogues of Virasoro primary fields are the {\em Virasoro biprimary fields},
which are simultaneously Virasoro primary under each of the two commuting
Virasoro operators. These fields were called {\em bitensor fields} in the
original paper, and, although they were originally given only for the coset
constructions, their form is the same for all affine-Virasoro constructions.

In this paper, we combine the three elements
\begin{enumerate}
\item Virasoro biprimary fields \cite{h3} in biconformal field theory
\item KZ-type null states \cite{kz} of each affine-Virasoro construction
\item Factorization \cite{ht,do,h3,gep} to a K-conjugate pair of ordinary
conformal field theories
\end{enumerate}
to derive a hierarchy of non-linear Ward identities for affine-Virasoro
correlators (see eq.(8.2)). The Ward identities properly follow from the first
two elements, and become non-linear differential equations on the assumption
of factorization, which we argue is consistent at an abstract level.

The abstract form of the Ward identities is only a first step toward the
correlators, however, because solution of the equations requires specific
factorization ans\"atze, which may vary over affine-Virasoro space.

As a first example, we have solved the non-linear equations for the simplest
non-trivial K-conjugate pairs,  $h\subset g$ and the $g/h$ coset constructions,
using a matrix factorization. The resulting coset correlators solve first-order
linear partial differential equations, with flat connections, whose solutions
are the coset blocks defined by Douglas \cite{do}.

The Conclusion speculates on other possible factorization ans\"atze, and,
following our clue in the coset constructions, we speculate briefly about flat
connections for all affine-Virasoro constructions.

\section{The Virasoro Master Equation}
In this section, we review the Virasoro master equation and some features
of the system which will be useful below.

The general construction begins with the currents of untwisted affine $g$
\cite{km,bh}
$$ J_a(z) = \sum_m J_a^{(m)} z^{-m-1}\;\;\;\;,\;\;\;a =1, \ldots , {\rm dim}\,g
\;\; \;, \;\;\;\;m,n \in \Z \eqno(2.1a) $$
$$J_a(z)J_b(w)={G_{ab} \over (z-w)^2}+i{f_{ab}}^c \left( {1 \over z-w}+
\frac{1}{2} \partial_w \right) J_c(w) +T_{ab}(w) +{\cal O}(z-w) \eqno(2.1b) $$
where ${f_{ab}}^c$ and $G_{ab}$ are respectively the structure constants and
general Killing metric of $g$. The current algebra (2.1) is completely general
since $g$ is not necessarily compact or semisimple. In particular, to obtain
level $x_I = 2k_I/\psi_I^2$ of $g_I$ in $g = \oplus_I g_I$ with dual Coxeter
number $\tilde{h}_I=Q_I/\psi_I^2$, take
$$G_{ab}=\oplus_I k_I \m_{ab}^I\;\;\;\;,\;\;\;\;
{f_{ac}}^d {f_{bd}}^c = - \oplus_I Q_I \m_{ab}^I \eqno(2.2) $$
where  $\m_{ab}^I$ and $\psi_I$ are respectively the Killing metric and the
highest root of $g_I$.

Next, consider the class of operators quadratic in the currents
$$ T(z)= L^{ab} \xx J_a (z)J_b (z)\xx =\sum_m L^{(m)}z^{-m-2} \eqno(2.3) $$
where $ T_{ab} = \xx J_a J_b \xx =T_{ba}$ is the composite two-current operator
in (2.1b). The set of coefficients $L^{ab}= L^{ba}$ is called the inverse
inertia tensor, in analogy with the spinning top. The requirement that $T(z)$
is a Virasoro operator
$$T(z)T(w)={c/2 \over (z-w)^4} +\left(\frac{2}{(z-w)^2}+{ \pa_w \over z-w}
\right)T(w)+{\rm reg.}\eqno(2.4)$$
restricts the values of the inverse inertia tensor to those which solve the
Virasoro master equation \cite{hk,rus}
$$ L^{ab}=2 L^{ac}G_{cd}L^{db}-L^{cd}L^{ef}{f_{ce}}^{a}{f_{df}}^{b}-
L^{cd}{f_{ce}}^{f}{f_{df}}^{(a}L^{b)e} \eqno(2.5a)$$
$$ c=2 G_{ab}L^{ab}\;.\eqno(2.5b)$$
The Virasoro master equation has been identified \cite{gme} as an Einstein-like
system on the group manifold, with $L^{ab}$ the inverse metric on tangent space
and $c={\rm dim}\,g-4R$, where $R$ is the Einstein curvature scalar.

Some general features of the Virasoro master equation include:
\begin{enumerate}
\item The affine-Sugawara construction \cite{bh,h1,wit1,kz} $L_g$ is
$$ L_g^{ab}=\oplus_I {\eta_I^{ab} \over 2k_I +Q_I}\;\;\;\,,\;\;\;\;\;\;\;
c_g=\sum_I {x_I {\rm dim}\,g_I \over x_I + \tilde{h}_I } \eqno(2.6) $$
for arbitrary level of any $g$, and similarly for $L_h$ when $h \subset g$.
In what follows, we refer to the affine-Sugawara constructions as the A-S
constructions.
\item K-conjugation covariance \cite{bh,h1,go,hk}. When $L$ is a solution of
the
master equation on $g$, then so is the K-conjugate partner $\tilde{L}$ of $L$,
$$ \tilde{L}^{ab}=L_g^{ab}-L^{ab}\;\;\;,\;\;\;\;\;\;\;\tilde{c}=c_g-c
\eqno(2.7) $$
and the corresponding stress tensors form a K-conjugate pair of commuting
Virasoro algebras
$$ \tilde{T}(z) =\sum_m \tilde{L}^{(m)} z^{-m-2} \eqno(2.8a) $$
$$ \tilde{T}(z)\tilde{T}(w)={\tilde{c}/2 \over (z-w)^4}+\left(\frac{2}{(z-w)^2}
+\frac{1}{z-w} \partial_{w}\right)\tilde{T}(w)+{\rm reg.}\eqno(2.8b)$$
$$ T(z) \tilde{T} (w)= {\rm reg.} \;\;\;\;\;\; . \eqno(2.8c) $$
The affine-Virasoro stress tensors $T$ and $\tT$ are quasi-primary under the
A-S stress tensor $T_g=T + \tT$.

The simplest K-conjugate pairs are the subgroup constructions $L_h$ and the
corresponding $g/h$ coset constructions \cite{bh,h1,go}
$$ L_{g/h}^{ab} = L_g^{ab} - L_h^{ab} \;\;\;,\;\;\;\; c_{g/h} =c_g -c_h
\eqno(2.9)$$
while repeated K-conjugation on embedded subgroup sequences generates the
nested coset constructions \cite{wit2,nuc} and the affine-Virasoro nests
\cite{nuc}.
\item Non-chiral versions of affine-Virasoro constructions are formed as usual
by left-right doubling \cite{gva,rd}. Because the constructions come in
commuting K-conjugate pairs, the action formulation of the generic theory
\cite{gva} is a gauge theory, in which a given construction is gauged by the
Virasoro generators of its K-conjugate partner. The simplest gauge choice for
the physical Hilbert space of the generic theory $L$ is the set of states which
are Virasoro primary under the K-conjugate  construction
$$ \tilde{L}^{m>0} |L\;{\rm physical}\rangle =0 \eqno(2.10) $$
and vice versa for the $\tilde{L}$ theory. This is not a complete gauge fixing
for the coset constructions, nor for the more general affine $h$-invariant
conformal field theories \cite{lh}, all of which are special cases with some
residual affine Lie symmetry.
\end{enumerate}

\section{${\bf L}$-Bases and Virasoro Biprimary States}
In this section we review and extend the construction of the $L^{ab}$-broken
{\em biprimary states} \cite{h3,nuc}, which are Virasoro primary under both
$T$ and $\tilde{T}$. These states were called simultaneous affine-conformal
highest-weight (ACHW) states in the original work.

We begin with the vacuum of affine $g$
$$ J_a^{m \geq 0} |0\rangle = L^{m \geq -1} |0\rangle =
\tilde{L}^{m \geq -1} |0\rangle  =0 \eqno(3.1) $$
and introduce the affine primary field $R_g^I (\T)$
$$ J_a(z) R_g^I(\T,w) = \left( {1\over z-w} + { \partial_w \over 2 \D_g(\T) }
\right) R_g^J(\T,w) {(\T_a)_J}^I + (R_g)_a^I (\T,w) + {\cal
O}(z-w)\eqno(3.2a)$$
$$ J_a^{m \geq 0} R_g^I(\T,0) |0\rangle = \d_{m,0} R_g^J(\T,0) |0\rangle
{(\T_a)_J}^I \;\;\;,\;\;\;\; I,J=1,\ldots,{\rm dim}\,\T \eqno(3.2b) $$
which corresponds to matrix representation $\T$ of $g$. In the finite part of
(3.2a), $\D_g(\T)$ is the conformal weight of representation $\T$ under the A-S
construction $T_g$ and $(R_g)_a^I=\xx J_a R_g^I \xx$ is a composite field
defined by the OPE.

Consider the action of the affine-Virasoro construction $T=L^{ab}\xx J_a
J_b\xx$
on the affine primary states. It is easily verified with (3.2b) that
$$ L^{(0)} R_g^I(\T,0) |0\rangle \chi_I^{\a}(\T)  = \D_{\a}(\T)
R_g^I(\T,0) |0\rangle \chi_I^{\a}(\T)  \eqno(3.3a) $$
$$ L^{ab} {(\T_a \T_b)_I}^J \chi_J^{\a} (\T) = \D_{\a}(\T) \chi_I^{\a}(\T)
 \;\;\;,\;\;\;\;\a,\b=1,\ldots,{\rm dim}\,\T \eqno(3.3b) $$
where $\D_{\a}(\T)$ and $\chi_I^{\a}(\T)$ are the eigenvalues and eigenvectors
of the conformal weight matrix $\D(\T) = L^{ab} \T_a \T_b$. It is conventional
\cite{nuc} to call $\D_{\a}(\T)$ the $L^{ab}$-broken conformal weights of the
affine primary state, and it is convenient to work directly in the eigenbasis
of the conformal-weight matrix, which we call an $L$-{\em basis} of
representation  $\T$,
$$ R_g^{\a}(\T,z) \equiv R_g^I(\T,z) \chi_I^{\a} (\T) \eqno(3.4a) $$
$$ {(\T_a)_{\a}}^{\b} \equiv (\chi^* (\T))_{\a}^I {(\T_a)_I}^J \chi_J^{\b}(\T)
\eqno(3.4b) $$
$$ {\D(\T)_{\a}}^{\b} = L^{ab} {(\T_a \T_b)_{\a}}^{\b} =\d_{\a}^{\b}
\D_{\a}(\T)
\eqno(3.4c) $$
$$ J_a(z) R_g^{\a}(\T,w) = { R_g^{\b} (\T,w) \over z-w} {(\T_a)_{\b}}^{\a}+
{\rm reg.} \;\;\;\;\;\; . \eqno(3.4d) $$
In what follows, we refer to the eigenbasis of any relevant conformal weight
matrix as an $L$-basis. Including the higher modes of $T$ and $\tT$, one now
verifies that
$$ L^{m \geq 0} R_g^{\a} (\T,0) |0\rangle =\d_{m,0} \D_{\a}(\T)
R_g^{\a} (\T,0) |0\rangle \eqno(3.5a) $$
$$ \tilde{L}^{m \geq 0} R_g^{\a} (\T,0) |0\rangle =\d_{m,0} \tilde{\D}_{\a}(\T)
R_g^{\a} (\T,0) |0\rangle \eqno(3.5b) $$
$$ \D_{\a}(\T) + \tilde{\D}_{\a}(\T) = \D_{g}(\T) \;\;\;\;\;\; .\eqno(3.5c) $$
It is clear that, in an $L$-basis, the $L^{ab}$-broken affine primary states
are Virasoro {\em biprimary states} with conformal weights
$(\D_{\a}(\T),\tilde{\D}_{\a}(\T))$ under the K-conjugate partners $T$ and
$\tilde{T}$.

We also define the carrier-space metric $\m_{\a\b}(\T)$ in the $L$-basis of
$\T$, which is used to raise and lower indices. The inverse of the
carrier-space metric appears in the A-S two-point correlators
$$ \langle R_g^{\a} (\T^1,z) R_g^{\b}(\T^2,w) \rangle =
 {\m^{\a\b}(\T^1) \d(\T^2,\bar{\T}^1 ) \over (z-w)^{2\D_g(\T^1)} }\eqno(3.6a)
$$
$$ {(\bar{\T}_a)_{\a}}^{\b} = - \m_{\a\r} \m^{\b\s} {(\T_a)_{\s}}^{\r}
= -{(\T_a^{*})_{\a}}^{\b} \eqno(3.6b) $$
where the Kronecker-delta on the right of (3.6a) records that the correlators
vanish except when the second representation $\T^2$ is the complex conjugate
of the first, as defined in (3.6b). The two-point A-S correlators satisfy the
usual global Ward identity
$$ \langle R_g (\T^1,z) R_g(\T^2,w) \rangle (\T_a^1 + \T_a^2) =0
\;\;\;,\;\;\;\; a=1,\ldots,{\rm dim}\,g  \eqno(3.7) $$
which implies with (3.4c) that the carrier-space metric satisfies
$$ \eqalignno{ \m^{\a\b}(\T) (\D_{\a}(\T) -\D_{\b}(\T) ) & =
\m_{\a\b}(\T) (\D_{\a}(\T) -\D_{\b}(\T) ) = 0 &(3.8a) \cr
 \m^{\a\b}(\T) (\tilde{\D}_{\a}(\T) -\tilde{\D}_{\b}(\T) ) & =
\m_{\a\b}(\T) (\tilde{\D}_{\a}(\T) -\tilde{\D}_{\b}(\T) ) =0 &(3.8b) \cr}  $$
in each $L$-basis. These identities are trivial for the A-S constructions, but
will be useful below in the general case.

In the appropriate $L$-bases, affine secondary states are also Virasoro
biprimary. As an example, we give the results for the $L^{ab}$-broken
one-current states of the affine vacuum module
$$ \langle J_A(z) J_B (w) \rangle = {G_{AB} \over (z-w)^2 } \eqno(3.9a) $$
$$ L^{m \geq 0} J_A(0)|0\rangle =\d_{m,0} \D_A J_A(0)| 0\rangle \eqno(3.9b) $$
$$ \tilde{L}^{m \geq 0} J_A(0)|0\rangle =\d_{m,0} \tilde{\D}_A
J_A(0)| 0\rangle  \eqno(3.9c) $$
$$ \D_A + \tilde{\D}_A = 1 \eqno(3.9d) $$
$$ {M_A}^B(L) = 2G_{AC} L^{CB} + {f_{AC}}^E L^{CD} {f_{DE}}^B = \D_A \d_A^B
\eqno(3.9e) $$
$$ G_{AB}(\D_A-\D_B) = G^{AB}(\tD_A -\tD_B) =0 \eqno(3.9f) $$
where $A,B=1,\ldots,{\rm dim}\,g$ labels the currents in an $L$-basis of the
conformal weight matrix ${M_A}^B$. The identities (3.9f) follow in an $L$-basis
because ${M_a}^c G_{cb}$ is $a,b$ symmetric in any basis \cite{cf}.

\section{Virasoro Biprimary Fields}
Virasoro biprimary fields  were  first constructed in Ref. \cite{h3}, where
they
were called bitensor fields. We review and extend this development in the
language of OPE's, incorporating an observation due to Schrans \cite{sta}.

Let $\phi_g^{\a}(z)$ be a Virasoro primary field under the A-S construction on
$g$
$$ T_g(z) \phi_g^{\a}(w) =\left({\D_g \over (z-w)^2} +{\partial_w \over z-w}
\right) \phi_g^{\a}(w) + {\rm reg.} \eqno(4.1) $$
where an $L$-basis for $\f_g$ is assumed, so that $\f_g^{\a}(0)|0\rangle$ is
biprimary under $T$ and $\tT$. In what follows, we refer to $\{ \f_g^{\a} \}$
as the {\em A-S fields}, examples of which include the affine primary fields
and the currents
$$ \eqalignno{ T_g(z) R_g^{\a}(\T,w) &=\left({\D_g(\T) \over (z-w)^2}
+{\partial_w \over z-w} \right) R_g^{\a}(\T,w) + {\rm reg.} &(4.2a) \cr
 T_g(z) J_A(w) &=\left( {1 \over (z-w)^2} +{\partial_w \over z-w}
\right) J_A(w) + {\rm reg.} \;\;\;\;\;\;. &(4.2b) \cr} $$
It should be noted that, although (4.2a) is usually assumed \cite{kz} for the
affine primary fields, the form is strictly correct only for integer level of
affine compact $g$. This subtlety is discussed in Appendix A, which finds an
extra zero-norm operator contribution for non-unitary A-S constructions.

Because the affine-Virasoro stress tensors $T$ and $\tilde{T}$ are
quasi-primary
fields under $T_g$, we may infer quite generally that \cite{sta}
$$ T(z) \phi_g^{\a}(w) = {\D_{\a} \f_g^{\a}(w) \over (z-w)^2}+
{ \partial_w \phi_g^{\a}(w)+\df_g^{\a}(w) \over z-w} + {\rm reg.}\eqno(4.3a) $$
$$ \tilde{T}(z) \phi_g^{\a}(w) = {\tilde{\D}_{\a} \f_g^{\a}(w) \over (z-w)^2} +
{\partial_w \phi_g^{\a}(w)+\tdf_g^{\a}(w)  \over z-w} + {\rm reg.}\eqno(4.3b)$$
$$ \D_{\a}+\tilde{\D}_{\a} =\D_g \eqno(4.3c) $$
$$ \partial \phi_g^{\a} +\df_g^{\a} +\tdf_g^{\a}  =0 \eqno(4.3d) $$
where $(\D_{\a},\tD_{\a})$ are the $(T,\tT)$ conformal weights of the biprimary
states $\f_g^{\a}(0) |0\rangle$ and $(\df_g^{\a},\tdf_g^{\a})$ are extra terms
which are non-vanishing in the general case. As an example of the
characteristic OPE (4.3), we have the known OPE for the currents \cite{hk}
$$ T(z) J_A(w) = \D_A \left( {1 \over (z-w)^2} + {\pa_w \over z-w} \right)
J_A(w)+ {2 i L^{BC} {f_{BA}}^D T_{CD}(w) \over z-w }+ {\rm reg.} \eqno (4.4a)$$
$$ \delta J_A^g = (\D_A -1) \pa J_A + 2 i L^{BC} {f_{BA}}^D T_{CD}\eqno(4.4b)$$
where $T_{AB} = \xx J_A J_B \xx$ is the composite operator $T_{ab}$ in an
$L$-basis of the currents. The corresponding form of $\tilde{\delta} J_A^g$ is
obtained from (4.4b) with $\D \ra \tD$ and $L \ra \tL$. As another example, we
have computed
$$ T(z) R_g^{\a} (\T,w) = \D_{\a}(\T) \left( {1 \over (z-w)^2} +
{1 \over \D_g(\T)}  {\pa_w \over z-w} \right) R_g^{\a}(\T,w) $$
$$ + { 2 L^{ab} (R_g)_a^{\b} (\T,w) {(\T_b)_{\b}}^{\a}  \over z-w} +
{\rm reg. }  \eqno(4.5a) $$
$$ \delta R_g^{\a}=\left( {\D_{\a}(\T) \over \D_g(\T)} -1 \right)
\partial R_g^{\a}(\T)+2 L^{ab} (R_g)_a^{\b}(\T){(\T_b)_{\b}}^{\a} \eqno(4.5b)
$$
for the affine primary fields $R_g^{\a} (\T) $, where the composite operator
$(R_g)_a^{\a}(\T)= \xx J_a R_g^{\a} (\T) \xx $  is defined in eq.(3.2a).
Further details of this computation are given in Appendix A.

A number of results follow from the characteristic OPE (4.3) by standard
manipulations. First, we have the equivalent forms
$$ [ L^{(m)},\phi_g^{\a}(z) ]= z^m[z\partial_z \phi_g^{\a}(z) + (m+1) \D_{\a}
\phi_g^{\a} (z) + z \df_g^{\a}(z) ] \eqno(4.6a) $$
$$ [ \tilde{L}^{(m)},\phi_g^{\a}(z) ]= z^m[z\partial_z \phi_g^{\a}(z) +
(m+1) \tD_{\a} \phi_g^{\a}(z) + z \tdf_g^{\a}(z) ] \eqno(4.6b) $$
from which one recovers that $\phi_g^{\a}$ creates the Virasoro biprimary
states
$$ L^{m\geq 0} \phi_g^{\a} (0) |0\rangle
 = \d_{m,0} \D_{\a}\phi_g^{\a} (0) |0\rangle  \eqno(4.7a) $$
$$ \tilde{L}^{m\geq 0} \phi_g^{\a}(0) |0\rangle
 = \d_{m,0} \tilde{\D}_{\a} \phi_g^{\a} (0)  |0\rangle \eqno(4.7b) $$
as discussed in the previous section. Moreover, the relations
$$ \df_g^{\a}(z)=- [\tilde{L}^{(-1)},\phi_g^{\a}(z) ]\;\;\;,\;\;\;\;
\tdf_g^{\a}(z)=- [L^{(-1)},\phi_g^{\a}(z) ] \eqno(4.8) $$
also follow from (4.6) and (4.3d), so the extra terms $\df_g^{\a},\tdf_g^{\a}$
in (4.3) are directly linked to the existence of a non-trivial K-conjugate
theory. See also Section 7, where the extra terms in (4.3) are understood as a
consequence of factorization.

Finally, we recover from (4.6) the generalized stability conditions
\cite{bh,h3}
$$ [z^{-m}L^{(m)}-L^{(0)},\phi_g^{\a}(z)]=m\D_{\a} \phi_g^{\a} (z)\eqno(4.9a)$$
$$ [z^{-m} \tilde{L}^{(m)} - \tilde{L}^{(0)} , \phi_g^{\a}(z)] =
m \tilde{\D}_{\a} \phi_g^{\a} (z)  \eqno(4.9b) $$
which are independent of the extra terms, and whose form at $z=1$ was central
in the original construction of the biprimary fields.

We turn now to the Virasoro biprimary fields $\f^{\a}(\bz,z)$, which satisfy
$$ \eqalignno{ T(z) \phi^{\a}(\bar{w},w) &=\left(  {\D_{\a} \over (z-w)^2}
+{ \partial_w \over  z-w} \right) \phi^{\a}(\bar{w},w) + {\rm reg.} &(4.10a)
\cr
 \tilde{T}(\bz) \phi^{\a}(\bar{w},w)& =\left(  {\tilde{\D}_{\a} \over
(\bz-\bw)^2}  +{ \partial_{\bw}  \over  \bz-\bw} \right) \phi^{\a}(\bar{w},w)
+ {\rm reg.}  \;\;\;\;\;\; . &(4.10b) \cr} $$
These fields were called ``bitensor fields'' in the original paper \cite{h3},
and they have a number of equivalent forms
$$ \eqalignno{\phi^{\a} (\bz,z) &=z^{L^{(0)}}\bz^{\tilde{L}^{(0)}}\phi_g^{\a}
(1) z^{-L^{(0)}-\D_{\a}} \bz^{-\tilde{L}^{(0)}-\tD_{\a} }  &(4.11a) \cr
& =\left( {\bz \over z} \right)^{\tL^{(0)}} \phi_g^{\a}(z)
\left( {z \over \bz} \right)^{\tL^{(0)}+ \tD_{\a} } &(4.11b) \cr
& =\left( {z \over \bz} \right)^{L^{(0)}} \phi_g^{\a}(\bz)
\left( {\bz \over z} \right)^{L^{(0)}+\D_{\a} } &(4.11c) \cr
& ={\rm e}^{(\bz-z)\tL^{(-1)}} \phi_g^{\a}(z) {\rm e}^{(z-\bz)\tL^{(-1)}}
&(4.11d) \cr
& ={\rm e}^{(z-\bz) L^{(-1)}} \phi_g^{\a}(\bz) {\rm e}^{(\bz-z) L^{(-1)}}
&(4.11e) \cr} $$
each of which is an $SL(2)$ boost of the A-S field $\phi_g^{\a}$.

The first line (4.11a) is the original form of the biprimary fields, but the
equality of all the forms in (4.11) may be verified with the A-S boost
identities
$$ \phi_g^{\a} (z\bz ) = z^{L_g^{(0)} } \phi_g^{\a} (\bz)
z^{-L_g^{(0)}-\D_g } \eqno(4.12a) $$
$$ \phi_g^{\a} (z+ \bz) = {\rm e}^{zL_g^{(-1)} } \phi_g^{\a} (\bz)
{\rm e}^{-z L_g^{(-1)}} \eqno(4.12b) $$
and the boost identity
$$ \partial_z \left(  {\rm e}^{ (\bz-z) L^{(-1)}}
\left( {z \over \bz} \right)^{L^{(0)}} \phi_g^{\a}(\bz)
\left( {\bz \over z} \right)^{L^{(0)}+\D_{\a}}{\rm e}^{ (z-\bz) L^{(-1)}}
\right) =0 \eqno(4.13) $$
which itself follows from the stability condition (4.9a).

Following the original arguments \cite{h3}, a check of the form (4.11a) is
given in Appendix B. For the present discussion, the simplest check of the OPE
(4.10a) uses (4.3a), (4.8) and the form in (4.11d),
$$ \eqalign{
T(z) &\phi^{\a} (\bw,w) = {\rm e}^{(\bw-w)\tL^{(-1)} }  T(z) \phi_g^{\a}(w)
{\rm e}^{(w-\bw)\tL^{(-1)} } \cr
& = {\rm e}^{(\bw-w)\tL^{(-1)} }\left( {\D_{\a} \phi_g^{\a}(w) \over (z-w)^2 }
+  { \partial_w \phi_g^{\a}(w) - [\tL^{(-1)},\phi_g^{\a}(w)] \over z-w} +
{\rm reg.} \right) {\rm e}^{(w-\bw)\tL^{(-1)} } \cr
&= \left( {\D_{\a} \over (z-w)^2} +{\partial_w \over z-w} \right)
\phi^{\a} (\bw,w) + {\rm reg.} \;\;\;\;\;\;. \cr } \eqno(4.14) $$
Similarly, the simplest check of the OPE (4.10b) follows the same steps from
the form in (4.11e).

Other useful properties of the biprimary fields include
$$ \langle \phi^{\a} (\bz,z) \rangle =0 \eqno(4.15a) $$
$$  \phi^{\a}(z,z) = \phi_g^{\a} (z)  \eqno(4.15b) $$
$$ \partial_z \phi^{\a}(\bz,z)\ve_{\bz=z} = [ L^{(-1)},\phi_g^{\a}(z)]
\eqno(4.15c) $$
$$ \partial_{\bz} \phi^{\a}(\bz,z)\ve_{\bz=z} = [ \tL^{(-1)},\phi_g^{\a}(z)]
\eqno(4.15d) $$
$$ (\partial_z +\partial_{\bz} )\phi^{\a} (\bz,z)\ve_{\bz=z} = \pa_z
\phi_g^{\a} (z) \eqno(4.15e) $$
$$\pa_{\bz}^q  \pa_{z}^p \f^{\a} (\bz,z) \ve_{\bz= z}= ({\rm ad}\,\tL^{(-1)}
)^q
 ({\rm ad}\,L^{(-1)} )^p \f_g^{\a} (z) \eqno(4.15f) $$
$$\phi^{\a}(0,0)|0 \rangle = \phi_g^{\a}(0)|0\rangle = \mbox{biprimary state}
\;\;\;\;\;\; . \eqno(4.15g) $$
See also Section 7, where the $SL(2)$-boost form (4.11) of the biprimary
fields is understood by factorization.

\section{Biconformal Field Theory}
It is clear from the discussion above that K-conjugate pairs of affine-Virasoro
constructions naturally form {\em biconformal field theories}, whose chiral
form includes two commuting Virasoro algebras. This is the viewpoint emphasized
in the Direct Approach to Operator Conformal Constructions \cite{h3}, the
Generic Affine-Virasoro Action \cite{gva}, and the review in Ref. \cite{rd}.

In biconformal field theories, the biprimary fields $\phi^{\a}(\bz,z)$ have
conformal weights $(\D_{\a},\tD_{\a})$, $\D_{\a}+\tD_{\a} =\D_g$, under the
K-conjugate pair of commuting stress tensors $(T,\tT)$, $T +\tT =T_g$, and
create the biprimary states $\phi^{\a}(0,0) |0\rangle $ with the same conformal
weights. The biprimary fields are in 1-1 correspondence with the A-S fields
$\f_g^{\a}(z)=\f^{\a}(z,z)$, which are Virasoro primary under the A-S
construction. Similarly, the correlators of the biprimary fields, called
biconformal correlators, reduce to their corresponding A-S correlators when
$\bz_i=z_i$ for all points $i$ in the correlator. Biconformal secondary fields
have the form $\pa_z^m\pa_{\bz}^n\f^{\a}(\bz,z)$ and biconformal secondary
states can be obtained with negative modes of $T$ or $\tT$, as usual.

The reader may enjoy the structural analogue between the chiral biconformal
field theories and ordinary non-chiral (or closed string) conformal field
theories, which also have two commuting Virasoro algebras. Indeed, it is to
highlight this analogy that we have written the argument of the biprimary
fields as $(\bz,z)$, though it is not necessary here to think of $\bz$ as the
complex conjugate of $z$.

As examples in this paper, we will focus on the biprimary fields of the
$L^{ab}$-broken affine primary fields
$$ R^{\a}(\T,\bz,z) ={\rm e}^{(\bz-z) \tL^{(-1)} } R_g^{\a} (\T,z)
{\rm e}^{(z-\bz) \tL^{(-1)}} \eqno (5.1a) $$
$$ \D_{\a}(\T) + \tD_{\a}(\T) =\D_g(\T) \eqno(5.1b) $$
and the biprimary fields of  the $L^{ab}$-broken currents
$$ \J_A(\bz,z) =  {\rm e}^{(\bz-z) \tL^{(-1)} } J_A(z)
{\rm e}^{(z-\bz) \tL^{(-1)}} \eqno (5.2a) $$
$$ \D_A + \tD_A = 1 \;\;\;\;\;\;. \eqno(5.2b) $$
Two- and three-point biconformal correlators and leading-term OPE's of the
biprimary fields are easily determined from the principles above, for example
$$ \langle R^{\a}(\T^1,\bz,z)  R^{\b}(\T^2,\bw,w) \rangle =
{\m^{\a\b}(\T^1) \d(\T^2,\bar{\T}^1) \over (z-w)^{2\D_{\a}(\T^1)}
(\bz-\bw)^{2\tD_{\a}(\T^1)}} \eqno(5.3a) $$
$$ \langle \J_A(\bz,z)  \J_B(\bw,w) \rangle =
{G_{AB}  \over (z-w)^{2\D_A} (\bz-\bw)^{2\tD_A} } \;\;\;\;\;\;. \eqno(5.3b) $$
In both cases, the denominators are fixed first by $SL(2) \times SL(2)$
covariance and the numerators are then fixed by comparison with the
corresponding A-S correlators (3.6a) and (3.9a) at $\bz=z$ and $\bw=w$.
Recalling eqs.(3.8) and (3.9f) in the form
$$ \m^{\a\b}(\T) = 0 \;\;\; {\rm when} \;\;\; \D_{\a}(\T ) \neq \D_{\b}(\T)
\eqno(5.4a) $$
$$ G_{AB} =  0 \;\;\; {\rm when} \;\;\; \D_A \neq \D_B  \eqno(5.4b) $$
we see that the biconformal correlators (5.3) are non-zero only when the
$L^{ab}$-broken conformal weights of the two biprimary fields are equal.

Similarly, we have
$$ \langle R^{\a_1}(\T^1,\bz_1,z_1)  R^{\a_2}(\T^2,\bz_2,z_2)
R^{\a_3}(\T^3,\bz_3,z_3) \rangle = { Y_g^{\a_1 \a_2 \a_3} \over
z_{12}^{\g_{12}} z_{13}^{\g_{13}} z_{23}^{\g_{23}}
 \bz_{12}^{\bc_{12}} \bz_{13}^{\bc_{13}} \bz_{23}^{\bc_{23}} } \eqno(5.5) $$
where $z_{ij}=z_i-z_j,\;  \bz_{ij}=\bz_i -\bz_j$ and
$$ \g_{ij} = \D_{\a_i}(\T^i) + \D_{\a_j}(\T^j) -\D_{\a_k}(\T^k)\;\;\;,\;\;\;\;
\bc_{ij} = \tD_{\a_i}(\T^i) + \tD_{\a_j}(\T^j) -\tD_{\a_k}(\T^k)  \eqno(5.6a)
$$
$$ \g_{ij} + \bc_{ij} = \g_{ij}^g = \D_g(\T^i) + \D_g(\T^j) -\D_g (\T^k)
\;\;\;\;\;\; .\eqno(5.6b) $$
The coefficient $Y_g$ is the invariant A-S three-point correlator which
satisfies the usual global Ward identity
$$  Y_g^{\b} ( \sum_{i=1}^3 {\T_a^i)_{\b}}^{\a} =0 \;\;\;,\;\;\;\;
a=1 , \ldots, {\rm dim}\,g \eqno(5.7) $$
where $\a$ and $\b$ are a shorthand for the index sets $\a_1\a_2\a_3$ and
$\b_1\b_2\b_3$.

For the $L^{ab}$-broken currents, we obtain in the same way
$$ \eqalign{
\langle \J_A(\bz_1,z_1) & \J_B(\bz_2,z_2) \J_C(\bz_3,z_3) \rangle \cr
& = { i{f_{AB}}^D  G_{DC}  \over \bz_{12} \bz_{13} \bz_{23} }
\left( {\bz_{12} \over z_{12} } \right)^{\D_A+\D_B-\D_C}
\left( {\bz_{13} \over z_{13} } \right)^{\D_A+\D_C-\D_B}
\left( {\bz_{23} \over z_{23} } \right)^{\D_B+\D_C-\D_A} \cr}  \eqno(5.8) $$
because $\g_{ij}^g =\D^g =1$. Then we also have  the most singular terms of
the $L^{ab}$-broken bilocal current algebra
$$ \eqalign{
\J_A(\bz,z) \J_B(\bw,w) = &{ G_{AB} \over (z-w)^{2\D_A} (\bz-\bw)^{2\tD_A} }
\cr
&+  \sum_C { i{f_{AB}}^C \J_C(\bw,w) \over (z-w)^{\D_A+\D_B-\D_C}
(\bz-\bw)^{\tD_A+\tD_B-\tD_C} } + \ldots \cr}  \eqno(5.9) $$
which follows from (5.8), (5.3b) and (3.9f).  Similarly, we have
$$ \eqalign{
\langle \J_A & ( \bz_1,z_1)  R^{\a_2}(\T^2,\bz_2,z_2)
R^{\a_3}(\T^3,\bz_3,z_3) \rangle \cr
= & { (\T_A^2)^{\a_2\a_3} \d(\T^2,\bar{\T}^3) \over
\bz_{12} \bz_{13}  \bz_{23}^{2 \D_g(\T^2)-1}  }
\left( {\bz_{12} \over z_{12} } \right)^{\D_A+\D_{\a_2}-\D_{\a_3}}
\left( {\bz_{13} \over z_{13} } \right)^{\D_A+\D_{\a_3}-\D_{\a_2}}
\left( {\bz_{23} \over z_{23} } \right)^{\D_{\a_2}+\D_{\a_3}-\D_A} \cr}
\eqno(5.10a) $$
$$ \J_A(\bz,z) R^{\a}(\T,\bw,w) = \sum_{\b}  { R^{\b}(\T,\bw,w)
{(\T_A)_{\b}}^{\a} \over (z-w)^{\D_A+\D_{\a}-\D_{\b}}
(\bz-\bw)^{\tD_A+\tD_{\a}-\tD_{\b}} } + \ldots  \eqno(5.10b) $$
where $\D_{\a_i}=\D_{\a_i}(\T^i)$ and $\tD_{\a_i}=\tD_{\a_i}(\T^i)$.
In these bilocal OPE's, the corresponding biconformal correlators can be used
to determine the contributions of the biconformal secondaries
$\pa_z^m \pa_{\bz}^n \f^{\a}(\bz,z) $ on the right.

More generally, we have the $SL(2)\times SL(2)$ decomposition
$$ \langle \phi_1^{\a_1}(\bz_1,z_1) \ldots \phi_n^{\a_n}(\bz_n,z_n) \rangle
\equiv \Phi^{\a_1 \ldots \a_n} (\bz,z)
={ Y^{\a_1 \ldots \a_n} (\bu, u ) \over \prod_{i <j}
z_{ij}^{\g_{ij}(\a) }   \bz_{ij}^{\bc_{ij}(\a) }  } \eqno(5.11a) $$
$$ \sum_{j\neq i} \g_{ij}(\a)= 2\D_{\a_i} \;\;\;,\;\;\;\;
 \sum_ {j\neq i} \bc_{ij}(\a)= 2\tD_{\a_i} \eqno(5.11b) $$
$$ \g_{ij}(\a)+\bc_{ij}(\a) = \g_{ij}^g\;\;\;,\;\;\;\;
\sum_{j \neq i} \g_{ij}^g=2(\D_{\a_i} + \tD_{\a_i}) =2 \D_i^g \eqno(5.11c) $$
$$ Y^{\a_1 \ldots \a_n} ( u, u ) = Y_g^{\a_1 \ldots \a_n} ( u )
\eqno(5.11d) $$
where $\{ \bu \}$ and $\{ u\}$ are the sets of independent cross-ratios
constructed from $\{\bz_i \}$ and $\{ z_i \}$ respectively, and $Y_g^{\a}( u )$
is the invariant correlator of the corresponding A-S fields $\f_g^{\a}$. In
general, the A-S correlator $\Phi_g^{\a} = \Phi^{\a}(z,z)$ and the
corresponding invariant A-S correlator $Y_g^{\a}$ also satisfy a global Ward
identity, for example
$$ Y_g^{\b} (\sum_{i=1}^n {\T_a^i)_{\b}}^{\a} =0 \;\;\;,\;\;\;\;
a= 1, \ldots ,{\rm dim}\,g  \eqno(5.12) $$
when broken affine primary fields $\f^{\a}(\bz,z)=R^{\a}(\T,\bz,z)$ are chosen
for the correlator.

We finally note that, up to this point, our development applies as well to the
interacting bosonic models \cite{dhs,k,ks,gep}, which include the
affine-Virasoro constructions in principle, and may be more general. Because
they exhibit K-conjugation covariance, these models also form biconformal field
theories: The analogues of the A-S fields $\f_g$ are sums of vertex operators
which satisfy the characteristic OPE (4.3), and the corresponding biprimary
fields $\f(\bz,z)$ are the $SL(2)$ boosts (4.11) of the vertex operator sums.
In what follows, however, a more central role is played by the organization
of our constructions on affine Lie algebra.

\section{Ward Identities}
The form (4.11) of the biprimary fields indicates that the biconformal
correlators can be constructed as power series expansions in A-S correlators
about the A-S point
$$ \langle \phi_1^{\a_1}(\bz_1,z_1) \ldots \phi_n^{\a_n}(\bz_n,z_n)
\rangle \ve_{\bz=z} = \langle \phi_{g,1}^{\a_1}(z_1) \ldots \phi_{g,n}^{\a_n}
(z_n)  \rangle \eqno(6.1) $$
and we have organized this expansion as a sequence of computations of the
Knizhnik-Zamolodchikov \cite{kz} type, expressed in the language of OPE's. The
equivalent language of KZ-type null states is discussed in Section 8.

For example, we know
$$ \eqalign{ \pa_i \langle \phi_1^{\a_1}(\bz_1,z_1) & \ldots
\phi_n^{\a_n}(\bz_n,z_n) \rangle \cr
= &\oint_{z_i} {{\rm d}w_i \over 2\p i} \langle T(w_i)
\phi_1^{\a_1}(\bz_1,z_1) \ldots \phi_n^{\a_n}(\bz_n,z_n)  \rangle \cr
= &\oint_{z_i} {{\rm d}w_i \over 2\p i} \oint_{w_i} {{\rm d}\m_i \over 2\p i}
\; {1 \over \m_i-w_i} \langle L^{ab} J_a(\m_i)  J_b(w_i)
\phi_1^{\a_1}(\bz_1,z_1) \ldots \phi_n^{\a_n}(\bz_n,z_n)\rangle\cr}\eqno(6.2)$$
where $\pa_i = \pa_{z_i}$, and so, at the A-S point we have
$$\eqalign{ \pa_i \langle \phi_1^{\a_1}(\bz_1,z_1) & \ldots \phi_n^{\a_n}
(\bz_n,z_n) \rangle \ve_{\bz = z} \cr
& = \oint_{z_i} {{\rm d}w \over 2\p i} \oint_w {{\rm d}\m \over 2\p i}
\; {1 \over \m-w} \langle L^{ab} J_a(\m)  J_b(w)
\phi_{g,1}^{\a_1}(z_1) \ldots \phi_{g,n}^{\a_n}(z_n)  \rangle \cr} \eqno(6.3)$$
where the right side of (6.3) is an A-S correlator. Similarly, we have
$$ \eqalign{\pa_i \pa_j \langle \phi_1^{\a_1}(\bz_1,z_1) &
\ldots \phi_n^{\a_n}(\bz_n,z_n) \rangle \ve_{\bz = z} \cr
& = \oint_{z_i} {{\rm d}w_i \over 2\p i} \oint_{w_i} {{\rm d}\m_i \over 2\p i}
\; {1 \over \m_i-w_i} \oint_{z_j} {{\rm d}w_j \over 2\p i} \oint_{w_j}
{{\rm d}\m_j \over 2\p i}\; {1 \over \m_j-w_j} \cr  & \;\;\;\;\;\;
\cdot \langle L^{ab} J_a(\m_i)  J_b(w_i) L^{cd} J_c(\m_j)  J_c(w_j)
\phi_{g,1}^{\a_1}(z_1) \ldots \phi_{g,n}^{\a_n}(z_n)\rangle  \cr}  \eqno(6.4)$$
and so on uniformly for any number of  derivatives. To obtain derivatives with
respect to one or more barred variables, say $\pa_i \rightarrow \bar{\pa}_i
=\pa_{\bz_i} $, replace $ L \rightarrow \tL$ in the $i$th current bilinear
on the right.

The right sides of these equations can be evaluated by standard dispersive
methods, using only the singular terms of the OPE's of the currents with the
A-S fields $\f_g^{\a}$. In what follows, we call these relations the
{\em Ward identities} of the biconformal field theories.

It is clear that the simplest Ward  identities will be obtained for the
biprimary fields $\f^{\a}(\bz,z) = R^{\a}(\T,\bz,z)$ of the $L^{ab}$-broken
affine  primaries $\f_g^{\a}(z) = R_g^{\a}(\T,z)$: In this case, the simple
algebra of $J$ and $R_g$  in (3.4d) will guarantee that the right sides of
these identities are  proportional to the A-S correlators themselves, whereas
extra inhomogeneous terms are generally obtained for broken affine secondaries.
In what follows, we focus on the broken affine primaries, although we have
collected some of the corresponding results for $L^{ab}$-broken currents in
Appendix C.

We present our results for the $L^{ab}$-broken affine primaries in the
simplified notation
$$ A^{\a}(\bz,z) \equiv  \langle R^{\a_1}(\T^1,\bz_1,z_1) \ldots
R^{\a_n}(\T^n,\bz_n,z_n) \rangle \eqno(6.5a) $$
$$ A_g^{\a}(z)  = A^{\a}(z,z) \equiv \langle R_g^{\a_1}(\T^1,z_1) \ldots
R_g^{\a_n}(\T^n,z_n) \rangle \eqno(6.5b) $$
where the subscript $g$ labels the A-S correlator. In this notation, the Ward
identities take the form
$$   \bpa_{j_1} \ldots \bpa_{j_q} \pa_{i_1} \ldots \pa_{i_p} A^{\a} \ve_{\bz=z}
= A^{\b}_g {( W_{ \bj_1 \ldots \bj_q i_1 \ldots i_p } )_{\b}}^{\a} \eqno(6.6)$$
where $W_{ \bj_1 \ldots \bj_q i_1 \ldots i_p} $ is a hierarchy of
{\em affine-Virasoro connections}, defined for each K-conjugate pair of
affine-Virasoro constructions. By construction, the connections are symmetric
under exchange of any pair of indices, for example $W_{ji} =W_{ij}$ and
$W_{\bj i} = W_{i \bj}$.

We have  computed the first- and second-order connections, which correspond
to first- and second-order derivatives. The first-order identities  are
$$ \pa_i A^{\a}\ve_{\bz=z} = A^{\b}_g {(W_i)_{\b}}^{\a} \;\;\;, \;\;\;\;
 W_i = 2 L^{ab} \sum_{j\neq i}  { \T_a^i \T_b^j \over z_{ij} } \eqno(6.7a) $$
$$ \bpa_i A^{\a}\ve_{\bz=z} = A^{\b}_g {(W_{\bi})_{\b}}^{\a} \;\;\;, \;\;\;\;
 W_{\bi} = 2\tL^{ab} \sum_{j\neq i} {\T_a^i \T_b^j \over z_{ij} } \eqno(6.7b)
$$
where the first-order connections $W_i ,W_{\bi}$ have the form of the A-S
connection \cite{kz}
$$ W_i^g =2L_g^{ab}  \sum_{j \neq i}{ \T_a^i \T_b^j \over z_{ij} }\eqno(6.8) $$
which is also obtained from $W_i$ when $L=L_g$.

The second-order identities are
$$ \pa_i \pa_j A^{\a}\ve_{\bz =z}=A_g^{\b} {(W_{ij})_{\b}}^{\a}\;\;\;,\;\;\;\;
W_{ij} = \pa_i W_j + \frac{1}{2} (W_i,W_j)_+ +  E_{ij} \eqno(6.9a)$$
$$ \bpa_i \bpa_j A^{\a} \ve_{\bz =z} =A_g^{\b} {(W_{\bi\,\bj})_{\b}}^{\a}
\;\;\;,
\;\;\;\;W_{\bi\, \bj} = \pa_i W_{\bj} + \frac{1}{2} (W_{\bi},W_{\bj})_+
+  E_{\bi \,\bj} \eqno(6.9b)$$
$$ \bpa_i \pa_j  A^{\a} \ve_{\bz =z} =A_g^{\b} {(W_{\bi j})_{\b}}^{\a}\;\;\;,
\;\;\;\; W_{\bi j} = W_{\bi} W_j + E_{\bi j} \eqno(6.9c)$$
where the extra terms $E$ in the second-order connections are
$$ E_{ij}=\left\{ \begin{array}{ll}
  2i L^{da}L^{e(b} {f_{de}}^{c)} \left\{
{ \T_c^j \T_b^j \T_a^i \over z_{ij}^2 }  +
\sum_{k \neq i,j}  {\T_c^k \T_b^j \T_a^i \over z_{ij} z_{jk} }  \right\}
+( i \leftrightarrow j)  & ,\;\;i \neq j \\
 -2i L^{da}L^{e(b} {f_{de}}^{c)} \sum_{l \neq i} \left\{
{ \T_c^l \T_b^l \T_a^i   + \T_c^i \T_b^i \T_a^l \over z_{il}^2} +
\sum_{k \neq i,l}  {\T_c^k \T_b^l \T_a^i \over z_{il} z_{ik} }  \right\}
& ,\;\; i =j \end{array} \right. \eqno(6.10a)$$
$$ E_{\bi \, \bj} = E_{ij} \ve_{L \rightarrow \tL } \;\;\;\;\;\;
.\eqno(6.10b)$$
$$ E_{\bi j} = \left\{ \begin{array}{ll}
- 2i L^{da}L^{e(b} {f_{de}}^{c)} \left\{
{ \T_c^j \T_b^j \T_a^i + \T_c^i \T_b^i \T_a^j\over z_{ij}^2}
- 2 \sum_{k \neq i,j}  {\T_c^k \T_b^i \T_a^j \over z_{ij} z_{ik} } \right\}
 & ,\;\;i \neq j  \\
 2i L^{da}L^{e(b} {f_{de}}^{c)} \sum_{l \neq i} \left\{
{ \T_c^l \T_b^l \T_a^i + \T_c^i \T_b^i \T_a^l\over z_{il}^2}
+ \sum_{k \neq i,l}  {\T_c^k \T_b^l \T_a^i \over z_{il} z_{ik} } \right\}
 & ,\;\; i=j \;. \end{array} \right.  \eqno(6.10c)$$
We note that the Virasoro master equation (2.5a) was employed in this
computation to  obtain the terms $ \pa_i W_j$ and $\pa_i W_{\bj}$ in (6.9),
which are linear in $L$ and $\tL$. The results (6.10) may also be collected in
the simpler form
$$ E_{ij} = -\frac{1}{2} ( E_{\bi j} + E_{ \bj i} ) \;\;\;\;,
\;\;\;  E_{\bi \, \bj} = E_{ij} \ve_{L \rightarrow \tL } \;\;\;\;,\;\;\;
E_{ \bi i} = - \sum_{ j \neq i} E_{\bi j} \eqno(6.11)$$
where $E_{\bi j}, \; i \neq j$ is given in (6.10c).

Using the results (6.7), (6.9) and (6.11), it is not difficult to verify the
first two orders of the general sum rules
$$ \sum_{i_k} W_{ \bi_i \ldots \bi_q i_1 \, ..\, i_k \, .. \, i_p} =
 \sum_{i_k} W_{ \bi_i \, .. \bi_k \, ..\bi_q i_1 \ldots i_p} = 0 \eqno(6.12) $$
which follow from the $SL(2) \times SL(2)$ decomposition (5.11).

Many other consistency relations must also hold among the affine-Virasoro
connections because the Ward identities (6.6) can be used to express the
biconformal correlators (6.5) in two ways
$$ \eqalignno{ A^{\a}(\bz,z) &= A_g^{\b} (\bz) \sum_{p=0}^{\infty} {1 \over p!}
\sum_{i_1 \ldots i_p} (z_{i_1}-\bz_{i_1}) \ldots (z_{i_p}-\bz_{i_p})
{(W_{i_1 \ldots i_p} (\bz) )_{\b}}^{\a} &(6.13a) \cr
 & = A_g^{\b} (z) \sum_{q=0}^{\infty} {1 \over q!}
\sum_{j_1 \ldots j_q} (\bz_{j_1}-z_{j_1}) \ldots (\bz_{j_q}-z_{j_q})
{(W_{\bj_1 \ldots \bj_q} (z) )_{\b}}^{\a} \;. \;\;\; &(6.13b) \cr } $$
Then, re-expansion of (6.13a) about $\bz = z$ implies (by comparison with
(6.13b)) that the barred connections $W_{\bj_1 \ldots \bj_q}$ can be expressed
in terms of the unbarred connections $W_{i_1 \ldots i_p}$ and the A-S
connection $W_i^g$. In Section 8, we will discuss these relations as members
of a larger set of relations which is necessary for the consistency of
factorization.

\section{Factorization}
In this section, we begin to examine the consistency and implications of
factorization \cite{ht,do,h3,gep}, which separates the biconformal field
theories into their corresponding K-conjugate pairs of ordinary conformal
field theories. Intuitively, each K-conjugate pair is a pair of
``square-roots''
of the affine-Sugawara construction.

To begin, we return to the biprimary fields, for which we assume the
{\em abstract} factorization
$$ \phi^{\a} (\bz,z) = ( \tf (\bz) \phi (z) )^{\a} \eqno(7.1a) $$
$$ \phi_g^{\a} (z) = \phi^{\a}(z,z) =( \tf (z) \phi (z) )^{\a} \eqno(7.1b) $$
where $\phi(z)$ and $\tf (z) $ are the proper fields of the $L$ theory
and the $\tL$ theory respectively. More precisely, $\f$ and $\tf$ are
 $(\D_{\a},0)$ and $(0,\tD_{\a})$ Virasoro primary fields respectively under
the K-conjugate stress tensors $(T,\tT)$, where the zeroes mean that
$ T(z) \tf (w) ={\rm reg.}$ and $\tT (z) \f (w) ={\rm reg.} $
By {\em abstract} factorization, we mean that we do not yet specify any
particular group index assignment for $\f (z)$ or $\tf(\bz)$ separately, and
much of our effort below concerns the consistency of factorization in this
abstract form. Ultimately, we must also study concrete factorizations - or
{\em factorization ans\"atze} - which may vary over affine-Virasoro space. As
an example, we will see below that the matrix factorization
$$ \f^{\a}(\bz,z)=  \tf_{g/h}^{\b} (\bz) {(\f_h(z))_{\b}}^{\a} \eqno(7.2) $$
is selected by the $g/h$ coset constructions, but, more generally, it may be
necessary to consider other ans\"atze, such as the symmetric factorization
$$ \f^{\a}(\bz,z)= \sum_{\nu} \tf_{\nu}^{\a} (\bz) \f_{\nu}^{\a}(z)\eqno(7.3)
$$
where $\nu$ is a conformal-block index whose range is to be determined.

We note first that the abstract factorization (7.1b) of the general A-S field
$\f_g$ provides both a consistency check and a deeper understanding of the
characteristic OPE (4.3):
$$ \eqalign{\;\;\;\; T(z) \f_g^{\a} (w) & = ( \tf (w) T(z) \f(w) )^{\a} \cr
& = ( \tf (w) \left( {\D_{\a} \over (z-w)^2} + {\pa_w \over z-w } \right)
\f(w) )^{\a} + {\rm reg.} \cr
& = { \D_{\a} \over (z-w)^2 } \f_g^{\a} (w) +
{ \pa_w \f_g^{\a} (w) + \df_g^{\a} (w) \over z-w} + {\rm reg. } \cr }
\eqno(7.4a) $$
$$ \df_g^{\a}(w) = - ( \pa_w \tf (w) \f(w) )^{\a} = - [\tL^{(-1)},\f_g^{\a}(w)]
\;\;\;\;\;\; .\eqno(7.4b) $$
In the same way, the K-conjugate relations
 $$ \tdf_g^{\a} (w) = - (\tf (w) \pa_w \f(w) )^{\a} = - [L^{(-1)},\f_g^{\a}(w)]
\eqno(7.5) $$
are obtained for the characteristic OPE with $\tT (z)$, so the extra terms
$\df_g^{\a} ,\tdf_g^{\a}$ are required by factorization when the K-conjugate
theory is non-trivial. The argument of this paragraph was obtained with
D. Gepner.

Similarly, the $SL(2)$-boost forms (4.11) of the biprimary fields are
consistent with factorization \cite{h3}, and factorization of the A-S
field implies factorization of the biprimary field: Beginning with the
factorized A-S field in (7.1b), we have
$$ \eqalign{ \f^{\a}(\bz,z) &= {\rm e}^{(\bz-z)\tL^{(-1)}} \f_g^{\a} (z)
{\rm e}^{(z-\bz)\tL^{(-1)}} \cr
& = ( {\rm e}^{(\bz-z)\tL^{(-1)}} \tf(z) {\rm e}^{(z-\bz)\tL^{(-1)}}
\f(z))^{\a} \cr &= (\tf (\bz) \f(z) )^{\a} \cr} \eqno(7.6) $$
where a standard $SL(2)$ boost in the $\tL$ theory, analogous to (4.12b),
was employed in the last step.

\section{Factorized Ward Identities}
We consider the corresponding abstract factorization of the biconformal
correlators
$$ A^{\a} (\bz,z) = ( \bA (\bz) A(z) )^{\a} \eqno(8.1)  $$
where $A,\bA$ are the {\em affine-Virasoro correlators}, that is, the proper
correlators of the $L$ theory and the $\tL$ theory respectively. Then the
Ward identities (6.6) become non-linear differential equations for the
affine-Virasoro correlators
$$ (\pa_{j_1} \ldots \pa_{j_q} \bA \;\pa_{i_1} \ldots \pa_{i_p} A)^{\a}
= A^{\b}_g {( W_{ \bj_1 \ldots \bj_q i_1 \ldots i_p  }
)_{\b}}^{\a}\eqno(8.2a)$$
$$ A_g^{\a} (z) = ( \bA (z) A(z) )^{\a} \eqno(8.2b) $$
$$A_g^{\b} (\sum_{i=1}^n {\T_a^i)_{\b}}^{\a} =0 \;\;\;,\;\;\;\;
a=1,\ldots,{\rm dim}\,g \;\;\;\;\;\; . \eqno(8.2c) $$
So long as factorization is held at this abstract level, the form of these
{\em factorized Ward identities} is universal across all the conformal field
theories of affine-Virasoro space. In this and the following section, we study
the consistency of these general systems at the abstract level.

To familiarize the reader with these equations, we begin by discussing the
first-order system
$$ (\bA \; \pa_i A)^{\a} = A_g^{\b} {(W_i)_{\b}}^{\a} \eqno(8.3a) $$
$$ (  \pa_i \bA \; A)^{\a} = A_g^{\b} {(W_{\bi})_{\b}}^{\a} \eqno(8.3b) $$
whose  connections are given in (6.7). As a first exercise, note that the
first-order equations and the global Ward identity (8.2c) on $A_g$ guarantee
$SL(2)$ covariance of the factorized correlators
$$ (\bA \, \sum_i \pa_i A)^{\a} = 0 \;\;\;,\;\;\;\;
(\bA \, \sum_i (z_i \pa_i +\D_{\a_i}(\T^i) ) A)^{\a} = 0  $$
$$ (\bA \, \sum_i  (z_i^2 \pa_i +2 z_i \D_{\a_i}(\T^i) ) A)^{\a} =0
\eqno(8.4)$$
and similarly on $\bA$ with $\D_{\a_i}(\T^i) \ra \tD_{\a_i}(\T^i)$.
Verification of these identities follows essentially standard lines, e.g.
$$ \eqalign{ \;\;\;\;\;\; (\bA \, \sum_i z_i \pa_i A)^{\a} & =
 A_g^{\b} \sum_i \sum_{j \neq i} L^{ab} {(\T_a^j \T_b^i)_{\b}}^{\a} \cr
& = - A_g^{\b} \sum_i L^{ab} {(\T_a^i \T_b^i)_{\b}}^{\a}
 = -A_g^{\a} \sum_i \D_{\a_i}(\T^i) \cr} \eqno(8.5) $$
where we have recalled in the last step that the conformal-weight matrices
$L^{ab} \T_a^i \T_b^i$ are diagonal in their respective $L$-bases.

A more important feature of the first-order equations is that they imply
the KZ-equations \cite{kz}
$$ \pa_i A_g^{\a} = \pa_i (\bA \, A)^{\a} = A_g^{\b} {(W_i^g)_{\b}}^{\a}
\eqno(8.6a) $$
$$ W_i + W_{\bi} = W_i^g \eqno(8.6b) $$
where $W_i^g$ is the A-S connection in (6.8). We shall see that the relation
(8.6b) is the first of a hierarchy of {\em factorization relations} necessary
for the consistency of factorization.

Moving on, we consider the second-order system
 $$ (\bA \;\pa_i \pa_j A)^{\a} = A_g^{\b} {(W_{ij})_{\b}}^{\a}\;\;\;, \;\;\;\;
(  \pa_i \pa_j \bA \;A)^{\a} = A_g^{\b} {(W_{\bi\,\bj})_{\b}}^{\a}\eqno(8.7a)
$$
$$ (\pa_i \bA \; \pa_j A)^{\a} = A_g^{\b} {(W_{\bi j})_{\b}}^{\a} \eqno(8.7b)
$$
whose connections are given in (6.9). These equations give a non-trivial check
of factorization because differentiation of the first-order equations requires
a second set of factorization relations
$$\eqalignno{ W_{ji} + W_{j \bi } & = (\pa_i +W_i^g) W_j &(8.8a) \cr
 W_{\bj i} + W_{\bj\,\bi} & = (\pa_i +W_i^g) W_{\bj} &(8.8b) \cr} $$
among the first and second-order connections. With some algebra and the
identity
$$ [W_i , W_j] = E_{\bi j} -E_{\bj i} \eqno(8.9) $$
we have verified that the factorization relations (8.8) are satisfied by the
explicit forms (6.7), (6.9) of the connections.

More generally, let
$$ W_{\{n\} } =
\{ W_{\bi_1 \ldots \bi_q \, i_{q+1} \ldots i_n } \,| 0 \leq q \leq n \} \;\;\;,
\;\;\;\; W_0 \equiv 1 \eqno(8.10) $$
be the set of $n+1$ connection types of order $n$. Among the $n$th-order
connections, we find by differentiation exactly $n$ factorization relations
which may be summarized in the form
$$ W_{ \{ n-1 \} i_{n} } +  W_{ \{ n-1 \} \bi_{n} } =
(\pa_{i_n} + W_{i_n}^g) W_{ \{n-1 \} }  \;\;\;\;\;\; .\eqno(8.11)$$
The relations (8.6b) and (8.8) are included in this set when $n=1$ and 2
respectively.

In fact, we can prove from the biprimary fields that all the factorization
relations among all the connections are satisfied: For example, the
differentiation relation (4.15e) implies that the correlators satisfy
$$ \{(\pa_i+\bpa_i) A^{\a}(\bz,z) \} \ve_{\bz =z}=\pa_i
A_g^{\a}(z)\eqno(8.12)$$
and, taken with (6.7), this identity establishes the first factorization
relation (8.6b). Similarly, we find the generalization of (8.12)
$$ \eqalign{
\{ (\pa_{i_n} +\bpa_{i_n} &)( \bpa_{i_1} \ldots \bpa_{i_q} \pa_{i_{q+1}}
\ldots \pa_{i_{n-1}} ) A^{\a} (\bz,z) \} \ve_{\bz = z} \cr
& = \pa_{i_n} \{ ( \bpa_{i_1} \ldots \bpa_{i_q} \pa_{i_{q+1}}
\ldots \pa_{i_{n-1}} ) A^{\a} (\bz,z) \} \ve_{\bz = z} \cr}  \eqno(8.13) $$
by repeated differentiation of the biprimary fields. Taken with the defining
relations (6.6) of the general connection, these identities establish the
general factorization relations (8.11). We conclude that the factorized
Ward identities (8.2) are consistent at the abstract level.

One consequence of the factorization relations is that, given the A-S
connection, there is only one independent connection type at each order, so
that all the connections can be expressed in terms of a canonical set, say
$$ W_i^g\;\;,\;\, W_i\;\;,\; \, W_{ij}\;\;,\;\, W_{ijk} \;\;, \;\,
W_{ijkl}\;\;
 , \ldots \;\;\;\;\;\; .\eqno(8.14) $$
As an illustration, we have $W_{\bi} = W_i^g -W_i $ at first order and
$$ \eqalignno{ W_{\bi j} & = (\pa_i + W_i^g) W_j - W_{ij} &(8.15a) \cr
 W_{\bi \, \bj} &= (\pa_i + W_i^g) (W_j^g-W_j) -(\pa_j + W_j^g) W_i +
 W_{ij} &(8.15b) \cr}  $$
at second order. In this form, the $n$th-order connections involve up to $n-1$
powers of the A-S covariant derivative $(\pa_i +W_i^g)$. The expressions
obtained in this way for the barred connections $W_{\bj_1 \ldots \bj_q}$ are
the consistency relations which follow by Taylor expansion of (6.13a) into
(6.13b).

Another consequence of the factorization relations is that, given the KZ
equations, all the Ward identities are solved by solving only one equation at
each order. As an example, the {\em complete set} of Ward identities
$$ \pa_i A_g^{\a} = A_g^{\b} {(W_i^g)_{\b}}^{\a}\;\;\;,\;\;\;\;
(\bA \,\pa_{i_1} \ldots \pa_{i_p} A)^{\a} =
A_g^{\b} {(W_{i_1 \ldots i_p})_{\b}}^{\a} \;\;\;,\;\;\;\; W_0 =1 \eqno(8.16a)$$
$$A_g^{\b} (\sum_{i=1}^n {\T_a^i)_{\b}}^{\a} =0 \;\;\;,\;\;\;\;
a=1,\ldots,{\rm dim}\,g  \eqno(8.16b) $$
involves only the canonical set of connections in (8.14). Combining the
complete set (8.16) with eqs.(6.13a) and (8.2b), we find
$$  \eqalign{
A^{\a}(\bz,z) &= (\bA (\bz) A(\bz) )^{\b} \sum_{p=0}^{\infty} {1 \over p!}
\sum_{i_1 \ldots i_p} (z_{i_1}-\bz_{i_1}) \ldots (z_{i_p}-\bz_{i_p})
{(W_{i_1 \ldots i_p} (\bz) )_{\b}}^{\a}  \cr
& = (\bA (\bz) A(z) )^{\a} \cr}  \eqno(8.17) $$
so that, as we saw for the biprimary fields, factorization of the biconformal
correlators follows from factorization of the A-S correlators.

We finally emphasize that the factorized Ward identities (8.2) are natural
generalizations of the KZ equations (8.6). We have seen that the identities
imply the KZ equations, and, in an equivalent language, the identities follow
from KZ-type null states. As an example, the identities
$$ | \chi^{\a} \rangle = \tL^{(-1)} R_g^{\a} (\T,0) | 0 \rangle - 2 \tL^{ab}
J_a^{(-1)} R_g^{\b} (\T,0) | 0 \rangle {(\T_b)_{\b}}^{\a} = 0 \eqno(8.18a) $$
$$ \langle 0 |  R_g^{\a_1} (\T^1,z_1) \ldots  R_g^{\a_n} (\T^n,z_n)
| \chi^{\a} \rangle = 0 \eqno(8.18b) $$
$$ [J_a^{(-1)} , R_g^{\a} (\T,z)] = z^{-1} R_g^{\b} (\T,z){(T_a)_{\b}}^{\a}
\eqno(8.18c) $$
$$ [ \tL^{(-1)}, R_g^{\a} (\T,z) ] = \bar{\pa}_z R^{\a}(\T,\bz,z) \ve_{\bz =z}
= (\pa_z \bar{R}(\T,z)\, R(\T,z) )^{\a} \eqno(8.18d) $$
imply the first-order factorized Ward identity (8.3b), while the unfactorized
Ward identity (6.7b) is obtained by stopping one step short in (8.18d).
Similarly, the K-conjugate Ward identities (8.3a) and (6.7a) follow by
$\tL \ra L$.

To obtain the A-S correlators as solutions of the Ward identities, consider
the K-conjugate pair
$$ \tL = L_g \;\;\;,\;\;\;\; L=0 \;\;\;,\;\;\;\; W_{\bi} = W_i^g \;\;\;,
\;\;\;\; W_i = 0 \eqno(8.19) $$
in (8.18) and (8.3), and the concrete factorization
$$ A^{\a} (\bz,z) = A^{\b} (\bz) {A_{\b}}^{\a}  = A_g^{\a} (\bz) \eqno(8.20) $$
in (8.3), where ${A_{\b}}^{\a}= \d_{\b}^{\a}$ are the correlators of the
trivial theory. Then (8.3a) is trivially satisfied and (8.3b) is the KZ
equation
on $g$. In this simple case, we obtain all the higher-order connections
$$ W_{\bj_1 \ldots \bj_q} = (\pa_{j_1} + W_{j_1}^g ) \ldots
(\pa_{j_{q-1}} + W_{j_{q-1}}^g ) W_{j_q}^g \;\;\;,\;\;\;\; W_{\{q-1\} i_q }=0
\eqno(8.21) $$
by iteration of the KZ equations and comparison with the Ward identities in
(8.2). The induced connections (8.21) agree with the by-hand second-order
connections in (6.9) (because $L_g^{e(b}{f_{de}}^{c)}=0$), and satisfy the
factorization relations (8.11) on inspection.

\section{Invariant Equations for Four-Point Correlators}
As further checks on the factorized Ward identities (8.2), we have substituted
the two- and three-point correlators (5.3a) and (5.5) into the first- and
second-order equations (8.3) and (8.7). Using the global Ward identity (8.2c),
we find that the equations are satisfied identically before or after
factorization.

To study the four-point correlators at this level, we introduce the
$SL(2) \times SL(2)$ decomposition (5.11), which involves the invariant
correlators
$$ Y^{\a_1 \a_2 \a_3 \a_4} ( \bu, u ) = Y^{\a} (\bu, u ) =
(\bY(\bu) Y(u) )^{\a} \eqno(9.1a) $$
$$ u = { z_{12} z_{34} \over z_{14} z_{32} } \;\;\;, \;\;\;\;
\bu = { \bz_{12} \bz_{34} \over \bz_{14} \bz_{32} } \eqno(9.1b)$$
and we choose the KZ gauge $\g_{12}=\g_{13}=\bc_{12}=\bc_{13}=0$ for
simplicity.
With this decomposition, the factorized Ward identities (8.2) reduce to the
one-dimensional invariant differential equations
$$ (\pa^q \bY(u) \, \pa^p Y(u) )^{\a} = Y_g^{\b}(u)
 {( W_{\bq p}(u) )_{\b}}^{\a} \eqno(9.2a) $$
$$ Y_g^{\a} (u) = (\bY(u) Y(u) )^{\a} \eqno(9.2b) $$
$$ Y_g^{\b} ( \sum_{i=1}^4 \T_a^i{ )_{\b}}^{\a} =0 \;\;\;,\;\;\;\;
a=1,\ldots,{\rm dim}\,g \eqno(9.2c) $$
where $W_{\bq p}$ are the {\em invariant connections} of order $ q+p$.
The explicit form of  the invariant connections through second order
$$ W_{\bar{0}0} = 1 \eqno(9.3a) $$
$$ W_{\bar{0}1} = \frac{2}{u}  L^{ab} \T_a^1 \T_b^2
+ \frac{2}{u-1}  L^{ab} \T_a^1 \T_b^3 \;\;\;, \;\;\;\;
 W_{\bar{1}0} = \frac{2}{u}  \tL^{ab} \T_a^1 \T_b^2
+ \frac{2}{u-1}  \tL^{ab} \T_a^1 \T_b^3 \eqno(9.3b)$$
$$ W_{\bar{0}2} = \pa W_{\bar{0}1} + W_{\bar{0}1}^2 + E_{\bar{0}2} \;\;\;,
\;\;\;\;W_{\bar{2}0} = \pa W_{\bar{1}0} + W_{\bar{1}0}^2 + E_{\bar{2}0}
\eqno(9.3c) $$
$$ W_{\bar{1}1} = W_{\bar{1}0} W_{\bar{0}1} - E_{\bar{0}2} =
W_{\bar{0}1} W_{\bar{1}0} - E_{\bar{2}0} \eqno(9.3d) $$
$$ \eqalign{ \;\;\;\;\;
E_{\bar{0}2} = -2i  L^{da} & L^{e(b} {f_{de}}^{c)} \left\{ \frac{1}{u^2} [
\T_a^1 \T_b^2 \T_c^2 + \T_a^2 \T_b^1 \T_c^1]  \right. \cr
 & \left. + \frac{1}{(u-1)^2}[\T_a^1 \T_b^3 \T_c^3 + \T_a^3 \T_b^1 \T_c^1] +
\frac{2}{u(u-1)} \T_a^1 \T_b^2 \T_c^3 \right\} \cr} \eqno(9.3e)$$
$$ E_{\bar{2}0} = E_{\bar{0}2} \ve_{L \ra \tL}  \eqno(9.3f)$$
was obtained by using the global Ward identity (9.2c) in the quadratic form
$$ Y_g^{\b} (  \sum_{j\neq i}^4 {(L^{ab} \T_a^i \T_b^j)_{\b}}^{\a} +
 \d_{\b}^{\a} \D_{\a_i} (\T^i) )  =0
 \;\;\;,\;\;\;\; i=1,2,3,4  \eqno(9.4)$$
to eliminate the fourth representation $\T^4$.

The first-order invariant equations imply the invariant KZ equations
$$  \pa Y_g^{\a} = \pa (\bY \, Y)^{\a} = Y_g^{\b}
{(W^g)_{\b}}^{\a}\eqno(9.5a)$$
$$ W^g  = W_{\bar{0}1} +W_{\bar{1}0}= \frac{2}{u} L_g^{ab} \T_a^1 \T_b^2
+ \frac{2}{u-1}  L_g ^{ab} \T_a^1 \T_b^3   \eqno(9.5b)$$
as in the previous section, where (9.5b) is the first factorization relation
on the invariant connections. We have also checked that the second-order
factorization relations
$$ W_{\bar{0}2} + W_{\bar{1}1} = (\pa + W^g) W_{\bar{0}1} \eqno (9.6a) $$
$$ W_{\bar{2}0} + W_{\bar{1}1} = (\pa + W^g) W_{\bar{1}0} \eqno (9.6b) $$
are satisfied identically by the invariant connections (9.3). The general
factorization relations among the invariant connections
$$ W_{\bq +1,p} + W_{ \bq , p+1} = (\pa + W^g) W_{\bq p} \eqno(9.7) $$
include the special cases (9.5b) and (9.6), and are presumably guaranteed by
the general factorization relation (8.11) among the connections.

Given the factorization relations (9.7) and the invariant connection $W^g$,
the steps of Section 8 show that there is only one independent invariant
connection at each order, and all the invariant connections can be expressed
in terms of, say, the canonical set $W^g$ and $W_{\bar{0} p}$. Similarly,
given the invariant KZ equations, only one invariant Ward identity must be
solved at each order, so that, for example, the collection
$$ \pa Y_g^{\a} = Y_g^{\b} {(W^g)_{\b}}^{\a} \;\;\;,\;\;\;\;
(\bY \, \pa^p Y )^{\a}= Y_g^{\b} {(W_{\bar{0}p})_{\b}}^{\a} \;\;\;,\;\;\;\;
W_{\bar{0} 0} = 1 \eqno(9.8a) $$
$$ Y_g^{\b} ( \sum_{i=1}^4 \T_a^i{ )_{\b}}^{\a} =0 \;\;\;,\;\;\;\;
a=1,\ldots,{\rm dim}\,g \eqno(9.8b) $$
is a complete set of invariant Ward identities.

Finally, we follow the reasoning at the end of Section 8 to obtain the
invariant
A-S correlators and their associated invariant connections
$$ Y^{\a}(\bu,u) = Y^{\b}_g(\bu) {Y_{\b}}^{\a} =Y_g^{\a}(\bu) \eqno(9.9a) $$
$$ W_{\bar{0}\,p} = \d_{p,0} \;\;\;,\;\;\;\; W_{\bar{q}p}=\d_{p,0}
(\pa +W^g)^{q-1} W^g \;\;\;,\;\;q \geq 1 \eqno(9.9b) $$
as solutions of the invariant Ward identities (9.2), where
${Y_{\b}}^{\a}=\d_{\b}^{\a}$ are the invariant correlators of the trivial
theory.

\section{Matrix Factorization of ${\bf h}$ and ${\bf g/h}$}
As a first non-trivial example, we solve the new Ward identities for the
correlators of the coset constructions \cite{bh,h1,go}. We will see that the
coset correlators satisfy first-order linear differential equations whose
solutions are the coset blocks defined by Douglas \cite{do}.

To begin, we collect some special properties of $h \subset g$ and $g/h$, which
we choose as $T= T_h$ and $\tT = T_{g/h}$. First, the generators of $h$ commute
with $T_h$ and $T_{g/h}$, so the conformal weights of the broken affine
primary fields satisfy
$$ [ \T_a,\D_h(\T) ]=[\T_a ,\D_{g/h}(\T)]=0 \;\;\;,\;\;\;\;a\in h\eqno(10.1a)$$
$$ {(\T_a)_{\a}}^{\b} (\D_{\a}^h(\T) -\D_{\b}^h(\T) )
 ={(\T_a)_{\a}}^{\b} (\D_{\a}^{g/h}(\T) -\D_{\b}^{g/h}(\T) )= 0 \eqno(10.1b) $$
in an $L$-basis of $\T$ (see Section 3). Second, the biconformal fields
(5.1a) of the broken affine primaries satisfy
$$ J_a (z) R^{\a} (\T,\bw,w) = {R^{\b}(\T,\bw,w) \over z-w} {(\T_a)_{\b}}^{\a}
+ {\rm reg.} \;\;\;,\;\;\;\;a \in h \eqno(10.2) $$
because the currents of $h$ commute with $\tT = T_{g/h}$. Third, we will need
the explicit form of the $h$ and $g/h$ connections through second order
$$ W_i^h  \equiv W_i = 2 L_h^{ab} \sum_{j\neq i}  { \T_a^i \T_b^j \over z_{ij}
}
 \;\;\;, \;\;\;\;\;\; W_i^{g/h} \equiv W_{\bi} = 2 L_{g/h}^{ab}
 \sum_{j\neq i}  { \T_a^i \T_b^j \over z_{ij} } \eqno(10.3a) $$
$$ W_{ij} = (\pa_i + W_i^h) W_j^h \;\;\;,\;\;\;\; \;\;
W_{\bj i} = W_j^{g/h} W_i^h \eqno(10.3b)$$
$$ W_{\bj \, \bi}
 = \pa_i W_j^{g/h} + W_i^{g} W_j^{g/h}  - W_j^{g/h} W_i^h \eqno(10.3c)$$
which follow from (6.9), (6.10) and (8.8) because
$L_h^{da} L_h^{e(b} {f_{de}}^{c)} = E_{ij}=  0$. Finally, we will study the
matrix factorization
$$ R^{\a}(\T,\bz,z) = (\bR_{g/h} (\T,\bz) R_h (\T,z))^{\a} =
 \bR_{g/h}^{\b} (\T,\bz) {(R_h(\T,z))_{\b}}^{\a} \eqno(10.4a)$$
$$ A^{\a}(\bz,z) = (\bA_{g/h} (\bz) A_h (z))^{\a} =
 \bA_{g/h}^{\b} (\bz) {(A_h(z))_{\b}}^{\a}  \eqno(10.4b)$$
announced in Section 7, where $\bA_{g/h}^{\a}$ are the {\em coset correlators}.

The equations for $h$ and $g/h$ are then
$$ \pa_{j_1} \ldots \pa_{j_q} \bA_{g/h}^{\b}\;\pa_{i_1} \ldots \pa_{i_p}
{(A_h)_{\b}}^{\a}  = A^{\b}_g
{( W_{ \bj_1 \ldots \bj_q i_1 \ldots i_p  } )_{\b}}^{\a}  \eqno(10.5a)$$
$$ A_g^{\a} (z) = \bA_{g/h}^{\b} (z) {(A_h(z))_{\b}}^{\a} \eqno(10.5b) $$
$$A_g^{\b} (\sum_{i=1}^n {\T_a^i)_{\b}}^{\a} =0 \;\;\;, \;\;\;\;
a=1,\ldots,{\rm dim}\,g  \eqno(10.5c) $$
$$ \bA_{g/h}^{\b}(\bz)  {(A_h(z))_{\b}}^{\g} (\sum_{i=1}^n {\T_a^i)_{\g}}^{\a}
=0 \;\;\;, \;\;\;\; a=1,\ldots,{\rm dim}\,h  \eqno(10.5d) $$
where (10.5a-c) are the factorized Ward identities (8.2) and the extra
$h$-global Ward identity in (10.5d) is implied by (10.2).

To solve this system, we focus on the first-order equations in the form
$$  \bA_{g/h}^{\b} {( \pa_i A_h - A_h W_i^h)_{\b}}^{\a} = 0 \eqno(10.6a) $$
$$  \pa_i \bA_{g/h}^{\b} \,  {(A_h)_{\b}}^{\a} = \bA_{g/h}^{\b}
{(A_h)_{\b}}^{\g} {(W_i^{g/h})_{\g}}^{\a} \;\;\;\;\;\;\; . \eqno(10.6b) $$
The first of these equations is solved by taking $A_h$ to be the (invertible)
evolution operator of the $h$-connection
$$ \pa_i {(A_h)_{\b}}^{\a} = {(A_h)_{\b}}^{\g} {(W_i^h)_{\g}}^{\a}
 \;\;\;,\;\;\;\; \pa_i {(A_h^{-1})_{\b}}^{\a} =
- {(W_i^h)_{\b}}^{\g} {(A_h^{-1})_{\g}}^{\a} \eqno(10.7) $$
that is, the KZ equations of $h$ embedded in $g$. Then (10.5b) gives the
coset correlators
$$ \bA_{g/h}^{\a} = A_g^{\b} {(A_h^{-1})_{\b}}^{\a} \eqno(10.8)$$
which solve (10.6b) in the form
$$ \pa_i \bA_{g/h}^{\a} = \bA_{g/h}^{\b} {(W_i[g/h])_{\b}}^{\a} \eqno(10.9a)$$
$$  W_i[g/h] = A_h W_i^{g/h} A_h^{-1} \;\;\;\;\;\;\; .\eqno(10.9b) $$
In what follows, we refer to the first-order linear partial differential
equations (10.9) as the {\em coset equations}, and we will call $W_i[g/h]$ the
{\em dressed coset connections}. Moreover, we will often suppress the group
indices on our equations, as in (10.9b).

Before proceeding, we emphasize that the dressed coset connections are flat
connections
$$ F_{ij} (W[g/h] )= A_h \{ F_{ij} (W^g) - F_{ij} (W^h)  \} A_h^{-1} = 0
\eqno(10.10a) $$
$$ F_{ij} (W) \equiv \pa_i W_j - \pa_j W_i + [W_i,W_j] \eqno(10.10b) $$
as they must be, by construction from (10.8). On the other hand, the dressed
connections are complicated by the $h$-dressing, and, at least generically, do
not fall in the class of connections associated to the classical Yang-Baxter
equation.

To see this explicitly, we consider high-level expansion \cite{hl}
of the inverse inertia tensors on simple $g$, restricting the correlators to
``low-spin'' representations with $\D(\T^i)={\cal O}(k^{-1})$ at high-level.
Then, it is not difficult to obtain the first few terms of $W_i^{g/h},\,A_h$
and the dressed coset connections
$$ W_i[g/h] = \sum_{j \neq i} {r_{ij}(z) \over z_{ij}} =
W_i^{g/h} + {\cal O}(k^{-2}) \eqno(10.11a) $$
$$ \eqalign{ r_{ij} (z) &= {P_{g/h}^{ab} \over k} \T_a^i \T_b^j
+ \frac{1}{2k^2} (P_h^{ab} Q_h -\m^{ab} Q_g) \T_a^i \T_b^j \cr
-&\frac{i}{k^2} P_{g/h}^{ab} P_h^{cd} {f_{bc}}^e \left(
\sum_{s \neq i} \ln \left( {z_{is} \over z_{is}^{0}} \right) \T_d^s\T_e^i\T_a^j
+ \sum_{s \neq j} \ln \left( {z_{js} \over z_{js}^{0}}
\right)\T_a^i\T_e^j\T_d^s
\right) + {\cal O}(k^{-3}) \cr} \eqno(10.11b) $$
where we have chosen the simplest initial condition $A_h(z_0)=1$ for the
evolution operator and defined the projection operators $P_h^{ab}$ and
$P_{g/h}^{ab}$ onto $h$ and $g/h$. Since $r_{ij}\neq r_{ij}(z_{ij})$, the
classical Yang-Baxter equation is excluded.

Moving on, we consider our solution vis-\`a-vis the higher-order Ward
identities (10.5). The $h$-evolution equation (10.7) and the coset equations
(10.9) induce the all-order  connections
$$ \eqalignno{
 W_{ \bj_1 \ldots \bj_q i_1 \ldots i_p  } & = W^{g/h}_{j_1 \ldots j_q}
 W^{h}_{i_1 \ldots i_p} \equiv W_{\{ q\}}^{g/h} W_{\{p\}}^h  &(10.12a) \cr
W_{\{p\}i_{p+1}}^h & = (\pa_{i_{p+1}} + W_{i_{p+1}}^h ) W_{\{p\}}^h
&(10.12b) \cr
 W_{ \{q\}j_{q+1}}^{g/h} = \pa_{j_{q+1}} & W_{\{q\}}^{g/h}+
W_{j_{q+1}}^g W_{\{q\}}^{g/h} - W_{\{q\}}^{g/h} W_{j_{q+1}}^h &(10.12c)\cr} $$
by comparison with the Ward identities (10.5). At second order, the induced
connections (10.12) agree precisely with our by-hand connections in (10.3b,c),
so our solution solves the second-order Ward identities. At higher order, it
is not difficult to check that the induced connections (10.12) solve the
general factorization relations (8.11).

Next, we verify that our solution solves the $h$-global Ward identity (10.5d).
To see this, note first that the coset correlators satisfy an $h$-global
identity
$$ \bA_{g/h} \sum_i \T_a^i = A_g A_h^{-1} \sum_i \T_a^i =
A_g \sum_i \T_a^i A_h^{-1} = 0 \;\;\;,\;\;\;\;
a=1,\ldots,{\rm dim}\,h \eqno(10.13)$$
because $A_h$ and $A_h^{-1}$ commute with the generators of $h$, while
the A-S construction $A_g$ satisfies the $g$-global Ward identity (10.5c).
By the same reasoning, we have
$$ \bA_{g/h} (\bz) A_h (z) \sum_i \T_a^i = \bA_{g/h} (\bz) \sum_i \T_a^i A_h(z)
 = 0 \;\;\;,\;\;\;\; a=1,\ldots,{\rm dim}\,h \eqno(10.14) $$
as required by (10.5d).

We turn now to verify the $SL(2)$ covariance of our solution, starting with
$g/h$. The coset equations and the $g$-global identity (10.5c) guarantee the
$SL(2)$ covariance of the coset correlators,
$$  \sum_i \pa_i \bA_{g/h}^{\a} =
 \sum_i (z_i \pa_i +\D_{\a_i}^{g/h} ) \bA_{g/h}^{\a} =
 \sum_i  (z_i^2 \pa_i +2 z_i \D_{\a_i}^{g/h} ) \bA_{g/h}^{\a} = 0
\eqno(10.15)$$
but the last two identities require a special trick, e.g.
$$ \eqalign{ \;\;\;\;\; \sum_i z_i \pa_i \bA_{g/h}^{\a}  &
=- \sum_{\b,\g} A_g^{\b} { \sum_i (L_{g/h}^{ab} \T_a^i \T_b^i )_{\b}}^{\g}
{(A_h^{-1})_{\g}}^{\a} \cr
 &= -  \sum_{\b} A_g^{\b}  \sum_i \D_{\b_i}^{g/h}
{(A_h^{-1})_{\b}}^{\a} =
-  \bA_{g/h}^{\a}  \sum_i \D_{\a_i}^{g/h} \;\;\;\;\;\;. \cr}  \eqno(10.16)$$
In the last step, we used one of the $\D$-exchange identities
$${(A_h)_{\a}}^{\b} (\D_{\a_i}^{g/h}(\T^i) -\D_{\b_i}^{g/h}(\T^i) ) =
 {(A_h)_{\a}}^{\b} (\D_{\a_i}^{h}(\T^i) -\D_{\b_i}^{h}(\T^i) )=0\eqno(10.17a)$$
$${(A_h^{-1})_{\a}}^{\b} (\D_{\a_i}^{g/h}(\T^i) -\D_{\b_i}^{g/h}(\T^i) ) =
 {(A_h^{-1})_{\a}}^{\b}
(\D_{\a_i}^{h}(\T^i) -\D_{\b_i}^{h}(\T^i) ) = 0 \eqno(10.17b)$$
which follow from (10.1) because $A_h$ and $A_h^{-1}$ are matrix-valued
functions of $\{ \T_a^i,\,a \in h\}$.

It is instructive to compare the statement of $SL(2)$ covariance in (10.15)
with the general statement (8.4) in this case, e.g.
$$( \sum_i (z_i \pa_i +\tD_{\a_i} )  \bA \, A)^{\a} =
    \sum_{\b} \sum_i (z_i \pa_i +\D_{\a_i}^{g/h} )  \bA_{g/h}^{\b}
{(A_h)_{\b}}^{\a} = 0  \;\;\;\;\;\;.\eqno(10.18)$$
This relation is verified from the coset equations by stopping one step short
in (10.16) and multiplying by $A_h$; equivalently, the two forms in (10.15) and
(10.18) differ by a $\D$-exchange identity. It is clear that the matrix
factorization (10.4) of $h$ and $g/h$ introduced an $\a_i \lra \b_i$ mismatch
in the general identity (10.18), which is corrected by the $\D$-exchange. Since
the $\D$-exchange identities are special to $h$ and $g/h$, it seems unlikely
that matrix factorization will suffice in the general case.

To see the $SL(2)$ covariance of the $h$ theory, we use the $h$-invariance
(10.13) of the coset correlators to rewrite the biconformal correlators as
$$  \bA_{g/h}^{\b}(\bz) {(A_h (z) )_{\b}}^{\a}
=  \bA_{g/h}^M (\bz)  {(A_h(z))_M}^{\a}  \eqno(10.19a)$$
$$\bA_{g/h}^{\a} =  v_M^{\a}(h) \bA_{g/h}^M  \;\;\;,\;\;\;\;\;
{(A_h)_M}^{\a} =  v_M^{\b} (h) {(A_h)_{\b}}^{\a} \eqno(10.19b)$$
$$  v_M^{\b}(h) {(\sum_i \T_a^i)_{\b}}^{\a} =0\;\;\;,\;\;\;\;
 a=1,\ldots,{\rm dim}\,h \;\;\;,\;\;\;\; M = 1, \ldots, {\rm dim}\,v(\T,h)
\eqno(10.19c)$$
where $ v_M^{\a_1 \ldots \a_n }(h) $ in (10.19c) are the $h$-invariant
tensors of $\T^1 \otimes \ldots \otimes \T^n$. The projections ${(A_h)_M}^{\a}$
of the evolution operator are the physical $h$-correlators because they satisfy
both the KZ equations of $h$ and an $h$-global Ward identity,
$$  \pa_i {(A_h)_M}^{\a} = {(A_h)_M}^{\b} {(W_i^h)_{\b}}^{\a}
\;\;\;,\;\;\; \; {(A_h)_M}^{\b} {(\sum_i \T_a^i)_{\b}}^{\a} = 0\;\;\;,\;\;\;\;
a=1,\ldots,{\rm dim}\,h  \eqno(10.20) $$
from which the usual $SL(2)$ identities are inferred.

\section{Coset Blocks}
To study the invariant four-point coset correlators, we introduce the
$SL(2)$ decompositions
$$ {(A_h(z))_M}^{\a} = { {(Y_h(u) )_M}^{\a} \over \prod_{i<j}^4
z_{ij}^{\g_{ij}(\a)} } \;\;\;, \;\;\;\;\;
 \bA_{g/h}^{\a}(z) = { \bY_{g/h}^{\a} (u) \over
\prod_{i<j}^4 z_{ij}^{\bc_{ij}(\a)} } \eqno(11.1a)$$
$$ {(Y_h)_M}^{\b}  (\sum_{i=1}^4 \T_a^i{)_{\b}}^{\a} =
\bY_{g/h}^{\b} (\sum_{i=1}^4 \T_a^i {)_{\b}}^{\a} = 0 \;\;\;,
\;\;\;\; a=1,\ldots,{\rm dim}\,h  \eqno(11.1b) $$
where the $h$-global Ward identities on the invariant correlators follow
from (10.20) and (10.13) by $\D$-exchange identities. This corresponds to the
invariant factorization
$$ Y^{\a}( \bu ,u) = (\bY_{g/h} (\bu) Y_h (u) )^{\a} =
 \bY_{g/h}^{\b} (\bu) {(Y_h(u))_{\b}}^{\a}
= \C^M(\bu) {(Y_h(u))_M}^{\a} \eqno(11.2a) $$
$$ {(Y_h)_M}^{\a} = v_M^{\b} (h) {(Y_h)_{\b}}^{\a}  \;\;\;,\;\;\;\;
\bY_{g/h}^{\a} = v_M^{\a}(h) \C^M   \eqno(11.2b) $$
and hence the factorized invariant Ward identities
$$ \pa^q \bY_{g/h}^{\b} \, \pa^p {(Y_h)_{\b}}^{\a} = Y_g^{\b}
{( W_{\bq p})_{\b}}^{\a}  \eqno(11.3a)  $$
$$ Y_g^{\a} = \bY_{g/h}^{\b} {(Y_h)_{\b}}^{\a} \eqno(11.3b) $$
$$ \;\;\;\;\;\; Y_g^{\b} (\sum_{i=1}^4 \T_a^i{)_{\b}}^{\a} =0
\;\;\;,\;\;\;\; a=1,\ldots,{\rm dim}\,g \;\;\;\;\;\; . \eqno(11.3c) $$
The invariant $h$ and $g/h$ connections through order two
$$ W^h \equiv  W_{\bar{0}1}  = \frac{2}{u}  L_h^{ab} \T_a^1 \T_b^2
+ \frac{2}{u-1}  L_h^{ab} \T_a^1 \T_b^3  \eqno(11.4a) $$
$$ W^{g/h} \equiv  W_{\bar{1}0}  = \frac{2}{u}  L_{g/h}^{ab} \T_a^1 \T_b^2
+ \frac{2}{u-1}   L_{g/h}^{ab} \T_a^1 \T_b^3  \eqno(11.4b) $$
$$  W_{\bar{0}2}  = (\pa + W^h) W^h \;\;\;,\;\;\;\;
 W_{\bar{1}1}  = W^{g/h} W^h \eqno(11.4c)$$
$$ W_{\bar{2}0}  = \pa W^{g/h} + W^g W^{g/h} - W^{g/h} W^h \eqno(11.4d)$$
also follow from Section 9 because
$L_h^{da} L_h^{e(b} {f_{de}}^{c)} =E_{\bar{0}2}=0 $.

At this point, the invariant solution
$$ \pa {(Y_h)_{\b}}^{\a} = {(Y_h)_{\b}}^{\g} {(W^h)_{\g}}^{\a} \;\;\; ,\;\;\;\;
\pa {(Y_h^{-1})_{\b}}^{\a} = - {(W^h)_{\b}}^{\g }{(Y_h^{-1})_{\g}}^{\a}
\eqno(11.5a) $$
$$ \bY_{g/h}^{\a} = Y_g^{\b} {(Y_h^{-1})_{\b}}^{\a} \eqno(11.5b)$$
$$ \pa  \bY_{g/h}^{\a} = \bY_{g/h}^{\b} {W[g/h]_{\b}}^{\a} \;\;\;, \;\;\;\;
 W[g/h] = Y_h W^{g/h} Y_h^{-1}  \eqno(11.5c) $$
can be inferred from (11.1) and the results of Section 10, or, equivalently,
from (11.3), following the arguments of Section 10. In what follows, we will
refer to (11.5b) as the {\em invariant coset correlators} and to (11.5c) as the
{\em invariant coset equations}. We also remark that the induced invariant
connections
$$ W_{\bq p} = W_q^{g/h} W_p^h  \eqno(11.6a)$$
$$ W_{p+1}^h = (\pa + W^h) W_p^h  \;\;\;,\;\;\;\;
 W_{q+1}^{g/h} = \pa W_q^{g/h} + W^g W_q^{g/h} - W_q^{g/h} W^h \eqno(11.6b)  $$
agree with the by-hand second-order invariant connections in (11.4c,d), and
satisfy the general factorization relations (9.7).

Two steps are needed to make the transition from the group basis to a more
familiar basis. First,  we introduce a $(g,h)$ invariant tensor basis
$$ \eqalignno{ Y_g^{\a}  & =   v_m^{\a} (g) \G^m \;\;\;,\;\;\;\;
 \bY_{g/h}^{\a} =  v_M^{\a}(h) \C^M   &(11.7a) \cr
 {(Y_h)_M}^{\a} & = v_M^{\b} (h) {(Y_h)_{\b}}^{\a} = {\H_M}^N v_N^{\a}(h)
 &(11.7b) \cr
 {(Y_h^{-1})_M}^{\a} & \equiv v_M^{\b} (h) {(Y_h^{-1})_{\b}}^{\a} =
  {(\H^{-1})_M}^N   v_N^{\a}(h) &(11.7c) \cr
 {(W^g)_m}^{\a}  & \equiv v_m^{\b}(g) {(W^g)_{\b}}^{\a} =
{(W^g)_m}^n v_n^{\a} (g) &(11.7d) \cr
 {(W^h)_M}^{\a} & \equiv v_M^{\b}(h) {(W^h)_{\b}}^{\a} =
{(W^h)_M}^N v_N^{\a} (h)  &(11.7e) \cr}  $$
where $\C^M$ are the invariant coset correlators in the tensor basis and
$ \{ v_m^{\a} (g) \} \subset \{ v_M^{\a}(h) \} $ are the $g$-invariant tensors
of $\T^1 \otimes \ldots \otimes \T^n$,
$$ v_m^{\b} (g) {( \sum_i \T_a^i)_{\b}}^{\a} = 0 \;\;\;, \;\;\;\;
a=1,\ldots,{\rm dim}\,g
\;\;\;,\;\;\;\; m =1 ,\ldots,{\rm dim}\,v(\T,g) \eqno(11.8) $$
which may be chosen to satisfy $v_m^{\a}(g) = v_m^{\a}(h)$. It follows that
$$ \C^M {\H_M}^N =  \G^n \d_n^N \eqno(11.9a) $$
$$ \C^M = \G^m {(\H^{-1})_m}^M \eqno(11.9b) $$
$$ \pa \G^m =  \G^n {(W^g)_n}^m \eqno(11.9c) $$
$$ \pa {\H_M}^N =  {\H_M}^L {(W^h)_L}^N \;\;\;, \;\;\;\;
 \pa {(\H^{-1})_M}^N = -  {(W^h)_M}^L {(\H^{-1})_L}^N  \eqno(11.9d) $$
and we will choose the initial condition
$$ {(\H(u_0))_M}^N = {(\H^{-1}(u_0))_M}^N = \d_M^N \eqno(11.10) $$
for the  evolution operators $\H$ and $\H^{-1}$ in (11.9d).

Second, we introduce a block basis,
$$ \G^m (u) = d^r {(\F_g (u))_r}^m  \eqno(11.11a) $$
$$ {(\H(u))_M}^N = {( \F_h^{-1} (u_0))_M}^R  {( \F_h (u))_R}^N \eqno(11.11b) $$
$$ {(\H^{-1}(u))_M}^N = {( \F_h^{-1} (u))_M}^R  {( \F_h (u_0))_R}^N
\eqno(11.11c) $$
where ${(\F_g)_r}^m$ and ${(\F_h)_R}^M$ are the usual conformal blocks of
$g$ and $h$, chosen so that the left indices $r$ and $R$ label the blocks by
$g$ and $h$  representations in the $s$ channel  ($u \ra 0$). Then
$$ \C^M (u) = d^r  {(\F_g(u))_r}^n {(\F_h^{-1}(u))_n}^R {(\F_h(u_0))_R}^M
\eqno(11.12)  $$
and we finally obtain
$$ \bY_{g/h}^{\a} (u) = d^r \, {\C(u)_r}^R w_R^{\a} (u_0,h) \eqno(11.13a) $$
$$ w_R^{\a}(u_0,h)  = {(\F_h(u_0))_R}^M v_M^{\a} (h) \eqno(11.13b) $$
$$ {\C(u)_r}^R =  {(\F_g(u))_r}^n   {(\F_h^{-1}(u))_n}^R  \eqno(11.13c)$$
$$ \pa {\C_r}^R = {\C_r}^S {W[g/h]_S}^R \eqno(11.13d) $$
$$ {W[g/h]_R}^S = {(\F_h)_R}^M [ \d_M^m {( W^g)_m}^l \d_l^L -
 {(W^h)_M}^L ] {(\F_h^{-1})_L}^S \eqno(11.13e) $$
where ${\C_r}^R$ in (11.13c) are the {\em coset blocks}, and (11.13d) are the
invariant coset equations in the block basis. With (10.19c) and (11.8), we
count
${\rm dim}\,v(\T,h) \cdot {\rm dim}\,v(\T,g)$ coset blocks for general
integrable representations $\{\T^i\}$ of $g$ and general $g/h$.

The coset blocks in (11.13c) are essentially those defined by Douglas
\cite{do},
who argued that they can be used to define consistent non-chiral conformal
field
theories.

\section{Simplification on ${\bf (g \times g)/ g }$}
Among the coset constructions and representations of $g$, we distinguish a
particularly simple class: For the cosets, we choose the type II symmetric
spaces
$$ {g \over h}  ={ \bg_{x_1} \times \bg_{x_2} \over
\bg_{x_1+x_2}}\eqno(12.1)$$
and  we consider the integrable representations of $\bg_{x_1}$
$$ {(\T_a^i)_{\a_i}}^{\b_i} = ({(\T_{\ba}^i)_{\ha_i}}^{\hb_i},0)
\;\;\;,\;\;\;\; \ba = 1 , \ldots ,{\rm dim}\,\bg  \;\;\;\;\;\; .\eqno(12.2)$$
In this case, we have the simplifications
$$ Y_g^{\a} (u) = Y_{\bg_{x_1}}^{\ha} (u) \;\;\;,\;\;\;\;
\G^M = d^R {(\F_{\bg_{x_1}} )_R}^M \;\;\;, \;\;\;\;
Y_h^{\a} (u) = Y_{\bg_{x_1+x_2}}^{\ha} (u) \eqno(12.3a) $$
$$ v_m^{\a}(g) = v_M^{\a}(h)=  v_M^{\ha} (\bg) \;\;\;,\;\;\;\;
m=M =1,\ldots,{\rm dim}\,v(\T,\bg) \eqno(12.3b)$$
$$ Y_{g/h}^{\a} (u) =Y_{g/h}^{\ha} (u) = d^R {\C(u)_R}^S w_S^{\ha}(u_0,\bg)
\eqno(12.3c) $$
$$ w_R^{\ha}(u_0,\bg)  = {(\F_{\bg_{x_1+x_2}}(u_0))_R}^M v_M^{\ha} (\bg)
\eqno(12.3d) $$
$$ {\C (u)_R}^S = {(\F_{\bg_{x_1}}(u))_R}^L
{(\F_{\bg_{x_1+x_2}}^{-1}(u))_L}^S  \eqno(12.3e)$$
where ${\rm dim}\,v(\T,\bg)$ is the number of $\bg$-invariant tensors and
$({\rm dim}\,v(\T,\bg))^2 $ is the number of coset blocks in the square
matrix $\C$. In (12.3), the blocks of $\bg_x$ satisfy
$$ \pa \F_{\bg_x} = \F_{\bg_x} W^{\bg_x} \;\;\;, \;\;\;\;
\pa \F_{\bg_x}^{-1} = -  W^{\bg_x} \F_{\bg_x}^{-1} \eqno(12.4a) $$
$$ W^{\bg_x} = \l_{\bg_x} \left({P \over u} + {Q \over u-1} \right)
\;\;\;,\;\;\;\; \l_{\bg_x} = {1 \over x + \hh_{\bg} }  \eqno(12.4b) $$
$$ v_M^{\hb}(\bg)  2\psi_{\bg}^{-2}
(\eta^{\ba\bb} \T_{\ba}^1 \T_{\bb}^2 {)_{\hb}}^{\ha}
=  {P_M}^N v_N^{\ha}(\bg) \eqno(12.4c)$$
$$ v_M^{\hb}(\bg)  2 \psi_{\bg}^{-2}
(\eta^{\ba\bb} \T_{\ba}^1 \T_{\bb}^3 {)_{\hb}}^{\ha}
= {Q_M}^N v_N^{\ha}(\bg) \eqno(12.4d)$$
where $\hh_{\bg}$ and $\psi_{\bg}$ are the dual Coxeter number and highest
root of $\bg$ respectively, and the square matrices
$P$ and $Q$ are defined in (12.4c,d). Finally, the iterated coset equations
$$ \pa^q \C = \C W_q[g/h] \;\;\;,\;\;\;\;
W_q[g/h] =  \F_{\bg_{x_1+x_2}} W_q^{g/h} \F_{\bg_{x_1 +x_2}}^{-1}
\eqno(12.5a)$$
$$ W_0^{g/h} =1 \;\;\;, \;\;\;\; W_1^{g/h} =(\l_{\bg_{x_1}}-\l_{\bg_{x_1+x_2}})
\left({P \over u} + {Q \over u-1} \right) \eqno(12.5b) $$
$$ \pa W_{q+1}^{g/h} = \pa W_q^{g/h} + W^{\bg_{x_1}} W_q^{g/h}
-W_q^{g/h} W^{\bg_{x_1+x_2}} \eqno(12.5c) $$
follow from (10.12c) and (12.4).

\section{An Example on ${\bf  (SU(n) \times SU(n) )/ SU(n) }$ }
As an example on $(\bg \times \bg)/\bg$, we specify
$(SU(n)_{x_1} \times SU(n)_{x_2})/SU(n)_{x_1+x_2}$ and we choose the
$\bg_{x_1} =SU(n)_{x_1}$ representations as
$$  \T^1=\T^4 = \T_{(n)} \;\;\;,\;\;\;\; \T^2=\T^3=\bar{\T}_{(n)} \eqno(13.1)$$
 where $\T_{(n)}$ is the fundamental representation of $SU(n)$ and
$\bar{\T}_{(n)}$ is its  complex conjugate, defined in (3.6b). These choices
correspond to the  correlators and conformal weights
$$ A_{\bg}^{\ha} =
\langle R_{\D_{\bg_x}}^{\ha_1} (\T,z_1) R_{\D_{\bg_x}}^{\ha_2} (\bar{\T},z_2)
R_{\D_{\bg_x}}^{\ha_3} (\bar{\T},z_3)R_{\D_{\bg_x}}^{\ha_4} (\T,z_4) \rangle
= {Y_{\bg}^{\ha} (u) \over (z_{14} z_{23})^{2 \D_{\bg_x}} } \eqno(13.2a) $$
$$ \D_{\bg_x}(\T_{(n)}) = \D_{\bg_x} = { n^2-1 \over 2n(x+n)} \eqno(13.2b) $$
$$ \bA_{g/h}^{\ha} =
\langle R_{\D_{g/h}}^{\ha_1} (\T,z_1) R_{\D_{g/h}}^{\ha_2} (\bar{\T},z_2)
R_{\D_{g/h}}^{\ha_3} (\bar{\T},z_3)R_{\D_{g/h}}^{\ha_4} (\T,z_4) \rangle
= {\bY_{g/h}^{\ha} (u) \over (z_{14} z_{23})^{2 \D_{g/h}} } \eqno(13.2c) $$
$$ \D_{g/h}(\T_{(n)}) = \D_{g/h} = \D_{\bg_{x_1}} -\D_{\bg_{x_1+x_2}} =
{x_2 (n^2-1) \over 2n (x_1+n) (x_1+x_2+n) }  \eqno(13.2d) $$
of $SU(n)_x$ and $g/h$ respectively. The $n$ conformal weights of each coset
field are degenerate in this case, so that any basis is an $L$-basis,
and we choose the Cartesian basis of Gell-Mann for simplicity.

In this basis, Knizhnik and Zamolodchikov \cite{kz} have provided us with the
required data on $SU(n)_x$,
$$ v_1^{\ha} (\bg) = \d^{\ha_1 \ha_2} \d^{\ha_3 \ha_4} \;\;\;,\; \;\;\;
v_2^{\ha}(\bg) = \d^{\ha_1 \ha_3} \d^{\ha_2 \ha_4} \;\;\;,\;\;\;\;
{\rm dim}\,v(\bg) =2   \eqno(13.3a)$$
$$ P =-\frac{1}{n} \left(  \matrix{ n^2-1 & 0 \cr n & -1} \right) \;\;\;,
\;\;\; \; Q =-\frac{1}{n} \left(  \matrix{ -1 & n \cr 0 & n^2-1} \right)
\;\;\;, \;\;\;\; \l_{\bg_x} = {1 \over x +n} \eqno(13.3b) $$
$$ \F_{\bg_x} = \left( \matrix{ {(\F_{\bg_x})_V}^1 &  {(\F_{\bg_x})_V}^2 \cr
               {(\F_{\bg_x})_A}^1 &  {(\F_{\bg_x})_A}^2 } \right)
\;\;\;,\;\;\;\; \D_{\bg_x}^A = {n \over x + n} \eqno(13.3c)$$
$$  \eqalign{
{(\F_{\bg_x}(u))_V}^1 & = u^{-2 \D_{\bg_x}} (1-u)^{\D_{\bg_x}^A - 2\D_{\bg_x}}
F(\lg,-\lg,1-n\lg;u) \cr
 {(\F_{\bg_x}(u))_A}^1  &= u^{ \D_{\bg_x}^A - 2\D_{\bg_x}}
(1-u)^{\D_{\bg_x}^A - 2\D_{\bg_x}} F((n-1)\lg,(n+1)\lg,1+ n\lg;u) \cr
{(\F_{\bg_x}(u))_V}^2  & = \frac{1}{x} u^{-2 \D_{\bg_x} +1}
(1-u)^{\D_{\bg_x}^A - 2\D_{\bg_x}}  F(1+\lg,1-\lg,2-n\lg;u) \cr
 {(\F_{\bg_x}(u))_A}^2  & =-n u^{ \D_{\bg_x}^A - 2\D_{\bg_x}}
(1-u)^{\D_{\bg_x}^A - 2\D_{\bg_x} }
F((n-1)\lg,(n+1)\lg,n\lg;u) \cr} \eqno(13.3d)$$
where $V,A$ label the vacuum and adjoint blocks in the $u \ra 0$ channel,
$\D_{\bg_x}^A$ is the conformal weight of the adjoint representation $\T_{(A)}$
and $F(a,b,c;u)$ is the hypergeometric function. We will also need the inverse
A-S blocks
$$ \F_{\bg_x}^{-1}  =
 -\frac{1}{n} (u(1-u))^{4\D_{\bg_x} - \D_{\bg_x}^A}
\left( \matrix{ {(\F_{\bg_x})_A}^2 &  -{(\F_{\bg_x})_V}^2 \cr
               - {(\F_{\bg_x})_A}^1 &  {(\F_{\bg_x})_V}^1 }
\right)\eqno(13.4)$$
and the crossing symmetry of the blocks \cite{kz}
$$  \F_{\bg_x} (u) =   X_{\bg_x} \F_{\bg_x} (1-u) \s_1  \;\;\;,\;\;\;\;
 \F_{\bg_x}^{-1} (u) =  \s_1 \F_{\bg_x}^{-1}  (1-u)  X_{\bg_x} \eqno(13.5a)  $$
$$ {(X_{\bg_x})_V}^V = n {\Ga( n \lg) \Ga (-n \lg) \over \Ga (\lg) \Ga (-\lg) }
\;\;\;,\;\;\;
{(X_{\bg_x})_A}^V = -n {\Ga( n \lg) ^2 \over \Ga ((n-1)\lg) \Ga ((n+1)\lg) }
\eqno(13.5b)$$
$$  {\rm Tr}\,X_{\bg_x} = 0 \;\;\;,\;\;\;\; {\rm det}\,X_{\bg_x} =-1
\;\;\;,\;\;\;\;  X_{\bg_x}^{-1} = X_{\bg_x} \eqno(13.5c)$$
where $\s_1$ is the first Pauli matrix and $\Ga$ is the gamma function.

The $({\rm dim}\,v(\T,\bg))^2=4$ coset blocks of the $g/h$ correlator in
(13.2c)
$$ \eqalign{ {\C(u)_V}^V &  =\frac{1}{n} u^{-2\D_{g/h} }
(1-u)^{-2\D_{g/h} + \D_{g/h}^{A,V} -1} \cr
 &[n F( \l_\bg, -  \l_\bg,1- n \l_\bg;u) F( (n-1)\l_h, (n+1) \l_h,  n \l_h;u)
\cr &  + {u \over  x_1} F(1+ \l_\bg, 1-  \l_\bg,2- n \l_\bg;u)
F( (n-1)\l_h, (n+1) \l_h,1+   n \l_h;u) ] \cr
{\C(u)_V}^A & = \frac{1}{n}
u^{-2\D_{g/h}+\D_{g/h}^{V,A} } (1-u)^{-2\D_{g/h} + \D_{g/h}^{A,V} -1} \cr
&  [ {1 \over x_1 + x_2}  F( \l_\bg,  - \l_\bg,1- n \l_\bg;u)
 F(1+ \l_h, 1-  \l_h,2- n \l_h;u)  \cr
&  - {1 \over x_1}  F(1+ \l_\bg,1 -  \l_\bg,2- n \l_\bg;u)
F( \l_h, -  \l_h,1- n \l_h;u) ] \cr
{\C(u)_A}^V  & =
 u^{-2\D_{g/h} + \D_{g/h}^{A,V}-1 } (1-u)^{-2\D_{g/h} + \D_{g/h}^{A,V} -1} \cr
& [  F( (n-1)\l_\bg,  (n+1) \l_\bg,1+ n \l_\bg;u) F( (n-1)\l_h, (n+1) \l_h,
  n \l_h;u)   \cr
&  - F( (n-1)\l_\bg,  (n+1) \l_\bg, n \l_\bg;u) F(( n-1)\l_h, (n+1) \l_h,
1+  n \l_h;u)  ] \cr
 {\C(u)_A}^A & = \frac{1}{n} u^{-2\D_{g/h} + \D_{g/h}^{A,A} }
(1-u)^{-2\D_{g/h} + \D_{g/h}^{A,V} -1} \cr
 & [ {u \over x_1 + x_2}  F( (n-1)\l_\bg,  (n+1) \l_\bg,1+ n \l_\bg;u)
F(1+ \l_h,1- \! \l_h, 2 - \! n \l_h;u)   \cr
&  + n F( (n-1)\l_\bg,  (n+1) \l_\bg, n \l_\bg;u) F( \l_h,-  \l_h,
1-  n \l_h;u)  ] \cr} \eqno(13.6a)$$
$$ \l_\bg \equiv \l_{\bg_{x_1}} = {1 \over x_1 +n}
 \;\;\;,\;\;\;\; \l_h \equiv \l_{\bg_{x_1+x_2}} = {1 \over x_1 +x_2 +n}
\eqno(13.6b) $$
$$ \eqalign{ \D_{g/h}^{V,V} & \equiv   0 \cr
 \D_{g/h}^{V,A} & \equiv 1 - \D_{\bg_{x_1+x_2}}^A =
{x_1 +x_2 \over x_1 + x_2+ n} \cr
 \D_{g/h}^{A,V} & \equiv \D_{\bg_{x_1}}^A +1 =
{x_1 +2n \over x_1 +n}  \cr
 \D_{g/h}^{A,A} & \equiv  \D_{\bg_{x_1}}^A - \D_{\bg_{x_1+x_2}}^A =
 { n x_2 \over (x_1 +n)(x_1 +x_2 +n) } \cr} \eqno(13.6c) $$
are now computed by substitution of (13.3) and (13.4) into the result (12.3e).
Note that
$$ {\C (u)_R}^S \mathrel{\mathop\sim_{u \ra 0}}
u^{-2 \D_{g/h}  + \D_{g/h}^{R,S}} \eqno(13.7) $$
so the results in (13.6c) are the conformal weights\footnote{The conformal
weights $\D_{g/h}^{V,V}$ and  $\D_{g/h}^{A,A}$ correspond to the
affine primary fields $(0,0,0)$ and $({\rm adjoint},0,{\rm adjoint})$, where
$( \T_{\bg_{x_1}}, \T_{\bg_{x_2}}, \T_{\bg_{x_1 +x_2}})$ denotes the branching
of $\T_{\bg_{x_1}} \otimes \T_{\bg_{x_2}}$ into representations of
$\bg_{x_1 + x_2}$. The fields of the other two conformal weights in (13.6c)
are apparently affine secondary in general.} of the $s$-channel coset blocks.

The crossing symmetry of the coset blocks
$$ \C (u) = X_{\bg_{x_1}} \C (1-u) X_{\bg_{x_1+x_2}} \eqno(13.8) $$
follows  from (13.5), so the coset blocks in the
 $t$ channel ($u \ra 1$) are the same as the $s$-channel. In the $u$  channel
($u \ra \infty$), we find four coset  blocks with conformal
weights\footnote{The fields with conformal weights $\D_{g/h}^{a,a}$ and
$\D_{g/h}^{s,s}$ are the affine primary fields $(a,0,a)$ and $(s,0,s)$
respectively, while the fields with $\D_{g/h}^{a,s}$ and $\D_{g/h}^{s,a}$
are apparently affine secondary in general.}
$$ \D_{g/h}^{a,a} \equiv \D_{\bg_{x_1}}^a - \D_{\bg_{x_1+x_2}}^a
\;\;\;,\;\;\;\;
\D_{g/h}^{a,s} \equiv \D_{\bg_{x_1}}^a - \D_{\bg_{x_1+x_2}}^s +1 \eqno(13.9a)
$$
$$ \D_{g/h}^{s,s} \equiv \D_{\bg_{x_1}}^s - \D_{\bg_{x_1+x_2}}^s
\;\;\;,\;\;\;\;
\D_{g/h}^{s,a} \equiv \D_{\bg_{x_1}}^s - \D_{\bg_{x_1+x_2}}^a +1 \eqno(13.9b)
$$
$$ \D_{\bg_x}^a= { (n-2)(n+1) \over n(x+n) }  \;\;\;,\;\;\;\;
\D_{\bg_x}^s={ (n+2)(n-1) \over n(x+n) }  \eqno(13.9c) $$
where $\D_{\bg_x}^{a}$ and $\D_{\bg_x}^s$ are the conformal weights of the
antisymmetric  and symmetric representations $\T_{(a)}$ and
$\T_{(s)}$ in $n \otimes n = (a) + (s) $. On $SU(2)_x$ and
$(SU(2)_{x_1} \times SU(2)_{x_2})/SU(2)_{x_1+x_2}$ the $u$-channel blocks are
the same as the $s$-channel blocks,
$$ R_{\D_{\bg_x}}^{\ha} ( \bar{\T}_{(2)}) \sim
R_{\D_{\bg_x}}^{\ha} ( \T_{(2)}) \;\;\;,\;\;\;\;
 R_{\D_{g/h}}^{\ha} ( \bar{\T}_{(2)}) \sim
R_{\D_{g/h}}^{\ha} ( \T_{(2)}) \eqno(13.10a) $$
$$ \Phi_{a,a} \sim \Phi_{V,V} \;\;\;,\;\;\;
\Phi_{a,s} \sim \Phi_{V,A} \;\;\;,\;\;\;
\Phi_{s,a} \sim \Phi_{A,V} \;\;\;,\;\;\;
\Phi_{s,s} \sim \Phi_{A,A} \eqno(13.10b) $$
$$ \D_{\bg_x}^s = \D_{\bg_x}^A \;\;\;, \;\;\;\; \D_{\bg_x}^a=0 \eqno(13.10c)$$
$$ \D_{g/h}^{a,a} = \D_{g/h}^{V,V} \;\;\;,\;\;\;
\D_{g/h}^{a,s} = \D_{g/h}^{V,A} \;\;\;,\;\;\;
\D_{g/h}^{s,a} = \D_{g/h}^{A,V} \;\;\;,\;\;\;
\D_{g/h}^{s,s} = \D_{g/h}^{A,A} \eqno(13.10d) $$
because $\T_{(2)} \sim \bar{\T}_{(2)}$, $\T_{(s)} = \T_{(A)}$ and
$\T_{(a)}=0$ for $SU(2)$.

Following the usual analyticity and crossing arguments \cite{bpz,kz}, we may
also construct the crossing-symmetric non-chiral  coset correlators
$$ \langle R_{\D_{g/h} } (z_1,z_1^*) \bar{R}_{\D_{g/h} } (z_2,z_2^*)
\bar{R}_{\D_{g/h} } (z_3,z_3^*) R_{\D_{g/h} } (z_4,z_4^*) \rangle =
{N \, Y(u,u^* ) \over | z_{14} z_{23} |^{4 \D_{g/h}} } \eqno(13.11a)$$
$$ \eqalign{ \;\;\;\; Y(u,u^* )=&  {\C(u)_V}^V  {\C(u^*)_V}^V
 + f(\l_h) {\C(u)_V}^A  {\C(u^*)_V}^A  \cr
 &+f(\l_\bg)^{-1} [ {\C(u)_A}^V {\C(u^*)_A}^V
 + f(\l_h) {\C(u)_A}^A {\C(u^*)_A}^A ] \cr}   \eqno(13.11b)$$
 $$ f(\a) \equiv  n^2 \left( \Ga (n\a) \over \Ga (1-n\a) \right)^2
{\Ga(1-(n-1)\a)\Ga (1-(n+1) \a) \over \Ga((n-1)\a) \Ga((n+1)\a)}\eqno(13.11c)
$$
where $N$ is a normalization and $z^*$, $u^*$ are the complex conjugates of
$z$, $u$.

{}From the non-chiral correlators we may infer the coset fusion rules for
$x_1 \neq 1$
$$ \eqalignno{
[\Fi ] \times [\bar{\Fi} ] &=[ \Fi_{V,V}] +[ \Fi_{V,A}] +[ \Fi_{A,V}] +
[\Fi_{A,A}  ] &(13.12a) \cr
[ \Fi ]\times [\Fi ] &= [\Fi_{a,a} ] + [ \Fi_{a,s}] + [\Fi_{s,a}] +[\Fi_{s,s}]
&(13.12b) \cr} $$
where $[\Fi ]$ and $[\bar{\Fi}]$ are the conformal blocks of $R_{g/h} (T)$ and
$R_{g/h} (\bar{T})$  respectively. At $x_1 =1$, however, only the coset blocks
${\C_V}^R$, $R=V,A$ contribute because
$$f( \l_{\bg_{x_1 = 1} })^{-1} = f \left( {1 \over n+1} \right)^{-1} = 0
\eqno(13.13) $$
and we obtain the truncated coset fusion rules
$$ \eqalignno{
[ \Fi ] \times [\bar{\Fi}] &=[ \Fi_{V,V} ] +[ \Fi_{V,A}] &(13.14a) \cr
 [\Fi ] \times [\Fi ] &= [\Fi_{a,a}] + [\Fi_{a,s} ] &(13.14b)\cr} $$
for $(SU(n)_1 \times SU(n)_{x_2})/SU(n)_{x_2+1}$. This consistent truncation
of the coset blocks is equivalent to a consistent affine cutoff on the blocks
of $g_{x_1} = SU(n)_{x_1}$,
$$ d^A = 0 \;\;\;, \;\;\;\; \G^M (u) = d^V {(\F_{\bg_{x_1= 1}}(u))_V}^M
\;\;\;, \;\;\;\; \bY_{g/h}^{\ha}(u) = d^V {\C(u)_V}^S w_S^{\ha} (u_0,\bg)
\eqno(13.15) $$
in which only the vacuum blocks ${(\F_{\bg_{x_1}})_V}^M$, $M=1,2$ of the A-S
construction are included at level one \cite{kz}.

Since the coset blocks (13.6) are unfamiliar and the non-chiral correlators
(13.11) are largely new, we have checked our results against the literature in
a number of cases.

Consider first the two-block solutions at $x_1 =1$, whose coset blocks
$$ \eqalign{ {\C (u)_V}^V  &=  u^{-2\D_{g/h} } (1-u) ^{-2\D_{g/h} }
 F((n+1)\l_h-1,(n-1) \l_h, n \l_h;u)  \cr
{ \C (u)_V}^A  & = {-x_2 \over n(x_2 +1)}  u^{-2\D_{g/h}+\D_{g/h}^{V,A} }
(1-u) ^{-2\D_{g/h}}  F(\l_h, 1-\l_h,2-n \l_h;u)  \cr} \eqno(13.16a)$$
$$ \D_{g/h} = { x_2 (n^2-1) \over 2n(n+ 1) (x_2 + n +1) } \;\;\;,\;\;\;\;
\D_{g/h}^{V,V} = 0 \;\;\; ,\;\;\;\;
\D_{g/h}^{V,A} = {x_2 + 1 \over x_2 + n +1} \eqno(13.16b) $$
satisfy the truncated fusion rules (13.14). The simple form of these blocks,
with only one hypergeometric function, is obtained from (13.6) because
$F(a,b,a;u) = (1-u)^{-b}$.

As a first check, the two-block solutions reduce to the correct results
\cite{bpz}
$$  R_{g/h}^{\ha} ( \T_{(2)}) \sim \Phi_{1,2} \eqno(13.17a)$$
$$\D_{g/h} = \D_{g/h} (\T_{(2)}) = \D_{1,2} = { m \over 4(m+3) }
\eqno(13.17b)$$
$$ \D_{g/h}^{V,V} = \D_{1,1} = 0   \;\;\;, \;\;\;\;
\D_{g/h}^{V,A} = \D_{1,3} =  {m+1 \over m+3}   \eqno(13.17c) $$
$$ [\Phi_{1,2}]  \times [ \Phi_{1,2}] =[\Phi_{1,1}]+[\Phi_{1,3}]\eqno(13.17d)$$
for the Virasoro minimal models $(SU(2)_1 \times SU(2)_m )/SU(2)_{m+1} $, where
the minimal model conformal weights $\D_{p,q}$ and the $SU(2)$ equivalences
(13.10) were used to make the  identifications. Moreover, the explicit form of
the coset blocks in (13.16) agrees with the blocks obtained by Dotsenko and
Fateev \cite{df} in this case. Finally, we have checked on
$(SU(3)_1 \times SU(3)_m)/SU(3)_{m+1}$ that the blocks in (13.16) are the known
\cite{w,miz} two-block solutions in the $W_3$ minimal models.

More generally, the iterated coset equations
$\pa^q {\C_V}^R = {\C_V}^S {W_q[g/h]_S}^R$ in (12.5) give the second-order
differential equation
$$ \eqalign{ & \left\{ {n+1 \over n [ 1 +{2 (n^2-1) \over n(n-1) } \D_{g/h} ] }
\pa^2 + \left[ {1 \over u} + {1 \over u-1} \right] \pa \right. \cr
& \left.  -\D_{g/h} \left[ \frac{2}{n}
{ [ 1 + {2 \over n-1} \D_{g/h}] \over [ 1 + {2(n^2-2) \over n(n-1) } \D_{g/h}]
}
\left( {1 \over u^2} + {1 \over (u-1)^2 } \right) -{ 2 \over u(u-1) } \right]
 \right\} {\C (u)_V}^R = 0  \cr}  \eqno(13.18)$$
for the entire set of two-block solutions, where $\D_{g/h}$ is given in
(13.16b). On $(SU(2)_1 \times SU(2)_m)/ SU(2)_{m+1}$,  this is the equation
derived by  BPZ from the chiral null state
$$ [(L_{g/h}^{(-1)})^2 -\frac{2}{3} (1 + 2  \D_{g/h} (\T_{(2)}))L_{g/h}^{(-2)}]
 R_{g/h}^{\ha} (\T_{(2)},0) | 0 \rangle = 0 \;\;,
\;\; R_{g/h}^{\ha} (\T_{(2)},0) | 0 \rangle = |\Phi_{1,2} \rangle\eqno(13.19)$$
where $ \D_{g/h} (\T_{(2)}) = \D_{1,2} $ is given in (13.17b).

As a check on the four-block solutions, we recover Douglas' conclusion for the
Virasoro minimal models
$$ R_{g/h}^{\ha} ( \T_{(2)}) \sim \Phi_{2,2}  \eqno(13.20a)$$
$$ \D_{g/h} = \D_{2,2}  = { 3 \over 4(m+2)(m+3) } \eqno(13.20b) $$
$$ \D_{g/h}^{V,V} = \D_{1,1} =  0  \;\;\;,\;\;\;\;
\D_{g/h}^{V,A} = \D_{1,3}  = {m+1 \over m+3}  \eqno(13.20c) $$
$$ \D_{g/h}^{A,V} = \D_{3,1}  = { m+4 \over m+2}   \;\;\;, \;\;\;\;
\D_{g/h}^{A,A} = \D_{3,3}   = {2 \over (m+2)(m+3)} \eqno(13.20d) $$
$$  [\Phi_{2,2}]  \times [ \Phi_{2,2}] =[\Phi_{1,1}] + [\Phi_{1,3}] +
[\Phi_{3,1}] + [\Phi_{3,3}] \eqno(13.20e)$$
when $n=2$, $x_1=m \neq 1$ and $x_2=1$. Moreover, the explicit form of the
coset blocks in (13.6) agrees with the blocks obtained by Zamolodchikov
\cite{z} in this case. Finally, we have checked that the four-block solutions
are in agreement with the known \cite{fqs,w} conformal weights and fusion rules
in the N=1 superconformal series ($ n=2$ and $x_1=2$, $x_2=m$ or
$x_1=m \neq 1 $, $x_2=2$) and $W_3$ minimal models ($n=3$ and
$x_1 =m \neq 1$, $x_2=1$).

For completeness,  we finally give the fourth-order differential equation of
the
general four-block solutions, obtained\footnote{The coefficients in (13.21)
were obtained by G. Rivlis using Mathematica.} from the iterated coset
equations in (12.5):
$$  \left\{ \pa^4 + a_1 \left[ {1 \over u} + {1 \over u-1} \right] \pa^3
 + \left[ a_2 \left( {1 \over u^2} + {1 \over (u-1)^2} \right)
 +{a_3 \over u(u-1) } \right] \pa^2 \right. $$
 $$ + \left[ a_4 \left( {1 \over u^3} + {1 \over (u-1)^3} \right)
 +a_5 \left( {1 \over u^2(u-1)} + {1 \over u(u-1)^2} \right) \right] \pa  $$
$$ + \left[ a_6 \left( {1 \over u^4} + {1 \over (u-1)^4} \right)
 +a_7 \left( {1 \over u^3(u-1)} + {1 \over u(u-1)^3} \right) \right. $$
 $$  \;\;\;\;\;\;\; \left. \left.
+{a_8 \over u^2(u-1)^2 }  \right] \right\}  {\C(u)_R}^S = 0
\;\;\;\;\;\;\;\;\;\;\;\;\; \eqno(13.21a) $$
$$a_1 = 4 \left[ 1 + { n^2-2 \over n^2 -1} \D_{g/h} \right] $$
$$ a_2 = 2 \left[ 1 + { 2(2n^2 -3) \over n^2 -1} \D_{g/h} +
{2 (n^4 -6n^2 +6) \over (n^2-1)^2 }\D_{g/h}^2 -
{n^2 \D_{g/h} \over c_{g/h} + 2n^2 \D_{g/h}} \right] $$
$$ a_3 =2 \left[ 5 +{ 10 (n^2-2) \over n^2-1 } \D_{g/h}
+{4(2n^4-7n^2 +6) \over (n^2-1)^2 } \D_{g/h}^2 + {2(n^2-2) \D_{g/h} \over
c_{g/h} +2n^2 \D_{g/h}} \right] $$
$$ a_4 = {2n^2 \D_{g/h} \over n^2-1} \left[
1 + {2(n^2-2) \over n^2-1} \D_{g/h}
-{8 (n^2-2) \over n^2(n^2-1)} \D_{g/h}^2  \right.  $$
$$ \left. -{ (n^2-1)+2(n^2-2) \D_{g/h} \over
c_{g/h} + 2n^2\D_{g/h}} \right] $$
$$ \eqalign{ a_5 = 2 & \left[ 1 + {5(n^2-2) \over n^2-1} \D_{g/h} +
{2(5n^4-20n^2+16) \over (n^2-1)^2} \D_{g/h}^2 \right. \cr
& \left. + {8(n^2-2)(n^2-3) \over (n^2-1)^2}
\D_{g/h}^3  + \left( { n^2-2 \over n^2-1 } \right)
{n^2-1 + 2(n^2-4) \D_{g/h} \over c_{g/h} +2n^2 \D_{g/h}} \D_{g/h} \right] \cr}
$$
$$ a_6= -{4 \D_{g/h}^2 \over n^2-1} \left[ 1 + {4 \over n^2-1}
\D_{g/h} - {4 \over n^2-1}
\D_{g/h}^2 - {2n^2 \D_{g/h} \over c_{g/h} +2n^2 \D_{g/h}}
\right]\eqno(13.21b)$$
$$ \eqalign{ a_7 = {4 \D_{g/h}^2 \over n^2-1}
& \left[ n^2 + {2n^2 (n^2-2) \over n^2-1} \D_{g/h}
- {8(n^2-2) \over n^2-1} \D_{g/h}^2 \right. \cr
& \left. - (n^2-2){(n^2-1) +2(n^2+2)\D_{g/h} \over c_{g/h} +2n^2\D_{g/h}}
\right] \cr} $$
$$ \eqalign{ a_8 = -{4 \D_{g/h}^2 \over n^2-1}&  \left[ (n^2+2) +8 \D_{g/h}
-{4(n^4-4n^2+6) \over n^2-1} \D_{g/h}^2  \right. \cr
 & \left. -{2(n^2-1)(n^2-2) +4(n^4-3n^2+4)\D_{g/h} \over c_{g/h} +2n^2\D_{g/h}}
 \right] \;\;\;\;\;\;\; . \cr} $$
Bearing in mind that we already have the solutions to this equation, its form
illustrates the technical superiority of the coset block approach over chiral
null state computations.

\section{Conclusion}

Following a review of the Virasoro master equation \cite{hk,rus} and Virasoro
biprimary fields \cite{h3} in biconformal field theory, we have derived a
hierarchy of non-linear Ward identities for affine-Virasoro correlators. The
hierarchy follows from KZ-type null states \cite{kz} and the assumption of
factorization \cite{ht,do,h3,gep}, whose consistency was verified at an
abstract level.

The abstract form of the Ward identities is only a first step toward the
correlators, however, because solution of the equations requires specific
factorization ans\"atze, which may vary over affine-Virasoro space.

In this paper, we solved the non-linear  equations only for the simple case of
$h\subset g$ and the  $g/h$ coset constructions \cite{bh,h1,go}, using a matrix
factorization: The resulting coset correlators satisfy first-order linear
partial differential equations, called the coset equations, whose solutions are
the coset blocks defined by Douglas \cite{do}. The coset equations exhibit a
class of flat connections, called the dressed coset connections, which are not
in the class of connections associated to the classical Yang-Baxter equation.

Beyond the coset constructions, we have noted in Sections 7 and 10 that other
factorization ans\"atze may be required, such as the symmetric factorization
$$ R^{\a} (\T,\bz, z) = (\bR(\T,\bz) R(\T,z))^{\a}  =
   \sum_{\nu} \bR_{\nu}^{\a} (\T,\bz) R_{\nu}^{\a} (\T,z)  \eqno(14.1a) $$
$$ A^{\a} (\bz, z) = ( \bA (\bz)  A (z) )^{\a} =
\sum_{\nu} \bA_{\nu}^{\a} (\bz) A_{\nu}^{\a} (z)  \eqno(14.1b) $$
$$ Y^{\a} (\bu, u) = ( \bY (\bu)  Y (u) )^{\a} =
\sum_{\nu} \bY_{\nu}^{\a} (\bu) Y_{\nu}^{\a} (u)  \eqno(14.1c) $$
where $\nu$ is a conformal-block index to be determined by the equations.

A more conservative direction keeps the matrix factorization and follows the
flat-connection clue provided by the coset constructions. By high-level
expansion \cite{hl} for low-spin representations on simple $g$, flat
connections
$W_i[L]$ can be associated to every high-level smooth \cite{gt} affine-Virasoro
construction $L$. The leading term of the flat connections
$$ W_i [L] = W_i + {\cal O}(k^{-2}) = 2L^{ab} \sum_{j\neq i}
{ \T_a^i \T_b^j \over z_{ij}} + {\cal O}(k^{-2}) \eqno(14.2a) $$
$$ L^{ab}= {P^{ab} \over 2k} + {\cal O}(k^{-2}) \eqno(14.2b) $$
is abelian-flat, where $P^{ab}$ is the high-level projection operator of the
construction. In the all-order expansion around (14.2), one finds a sequence
of abelian Bianchi identities which are satisfied so long as the leading term
is abelian-flat. These flat connections are correct for the A-S and coset
constructions, and, for all $L$, the corresponding high-level correlators
$$ A^{\a}[L] = v^{\b}(g) \left(\d_{\b}^{\a} + \frac{P^{ab}}{k} \sum_{i<j}
\ln \left({z_{ij}\over z_{ij}^0 }\right) {(\T_a^i \T_b^j)_{\b}}^{\a} \right)
+{\cal O}(k^{-2})\eqno(14.3) $$
satisfy $SL(2)$-covariance and the Ward identities at the indicated order. We
are presently investigating these properties at higher order.

We finally remark on an open question posed by Douglas for the coset
constructions. We saw in Section 13 that the set of coset blocks is sometimes
larger than the set of chiral blocks, as defined by chiral null-state
differential equations. The question posed by Douglas concerns the precise
relation between the two sets of blocks. Based on our examples, it is a
reasonable conjecture that the two sets are the same when the A-S blocks, and
hence the coset blocks, are restricted to the {\em integrable blocks},
$${ \C_{\hat{r}}}^{\hat{R}} ={( \F_g)_{\hat{r}}}^{n}
{(\F_h^{-1})_{n}}^{\hat{R}} \eqno(14.4) $$
where $\hat{r}$ and $\hat{R}$ are the integrable representations of $g$ and
$h$.
This truncation to the integrable blocks is always consistent (because affine
Lie algebra is a chiral construction), and ordinary differential equations for
the integrable blocks are obtainable in principle from the coset equations.
Without a better characterization of the set of all chiral blocks, however,
the conjecture seems difficult to prove or disprove.

\section*{Acknowledgements}
We acknowledge helpful conversations with O. Alvarez, M. Douglas, J.
Figueroa-O'Farrill, R. Flume, D. Gepner, A. Giveon, A. Le Clair, M. Porrati,
G. Rivlis, K. Schoutens, S. Schrans, K. Selivanov, A. Sevrin, M. Terhoeven and
W. Troost. Special thanks is due to E. Kiritsis, with whom we have shared a
long interest in this problem.
One of the authors (NO) gratefully
acknowledges the hospitality of the LBL theory group during part of this work.

\newpage
\bigskip \bigskip
\centerline{\bf Appendix A: Stress tensors and affine primary fields} \bigskip

\noindent The results of this appendix were obtained with E. Kiritsis.

We begin with the defining relations
$$ J_a(z)R_g^I(\T,w)=\! \left( {1 \over z-w} + {1 \over 2 \D_g(\T)}\pa_w
\right)
\! R_g^J(\T,w) {(\T_a)_J}^I + (R_g)_a^I (\T,w) + {\cal O}(z-w) \eqno(A.1a) $$
$$ \langle R_g^I (\T,z) R_g^J(\bar{\T},w) \rangle = { \m^{IJ} (\T)
 \over (z-w)^{2\D_g(\T)} } \eqno(A.1b) $$
for the affine primary field whose matrix representation is $\T$. $\bar{\T}$ is
the complex conjugate representation defined in (3.6b), and indices are raised
and lowered with the metric on carrier space $\m_{IJ}(\T)$, so that
$\bar{\T}=- \T^T= - \T^*$.

With (A.1), we may compute the correlators of any number of currents with two
primary fields $R_g(\T)R_g(\bar{\T})$. From these, we obtain the correlators
$$ \langle T_{ab}(z_1) R_g^I (\T,z_2) R_g^J (\bar{\T},z_3) \rangle
= { \frac{1}{2} (\T_a,\T_b)_+^{JI} \over z_{12}^2 z_{13}^2 z_{23}^{2 \D_g(\T)
-2}} \eqno(A.2a) $$
$$ \langle  (R_g)_a^I (\T,z) R_g^J (\bar{\T},w) \rangle = 0 \eqno(A.2b) $$
$$ \langle  (R_g)_a^I (\T,z) (R_g)_b^J (\bar{\T},w) \rangle
= {\left( G_{ab} \m (\T)+ {2 \D_g(\T) -1 \over 2\D_g(\T)} \T_b \T_a - \T_a \T_b
\right)^{JI} \over (z-w)^{2\D_g(\T) +2} } \eqno(A.2c) $$
$$ \eqalign{
 \langle & T_{ab}(z_1) R_g^I (\T,z_2) (R_g)_c^J (\bar{\T},z_3) \rangle \cr
&= { \left( G_{ac}\T_b + G_{bc} \T_a -(\T_a\T_c\T_b+\T_b\T_c\T_a) +
{2 \D_g(\T) -1 \over 2\D_g(\T)} \T_c (\T_a,\T_b)_+ \right)^{JI} \over
z_{12} z_{13}^3 z_{23}^{2 \D_g(\T) -1}  } \cr}  \eqno(A.2d) $$
where $T_{ab} = \xx J_a J_b \xx $ and $(R_g)_a^I(\T) = \xx J_a R_g^I(\T) \xx$.
These correlators determine  the OPE
$$ T_{ab}(z) R_g^I (\T,w) = \left( {1 \over (z-w)^2 }+ {1 \over \D_g(\T)}
{\pa_w \over z-w} \right) R_g^J(\T,w) \frac{1}{2}
{\left( (\T_a,\T_b)_+ \right)_J}^I $$
$$ + {1 \over z-w} (R_g)_{(a}^J (\T,w) {(\T_{b)})_J}^I + {\rm reg.
}\eqno(A.3)$$
which we take in an $L$-basis of representation $\T$ (see Section 3). Then,
multiplication by $L^{ab}$ gives the OPE of the affine-Virasoro contruction
$T =L^{ab} \xx J_a J_b \xx$ with the affine primary field
$$ T(z) R_g^{\a} (\T,w) = \D_{\a}(\T) \left( {1 \over (z-w)^2} +
{1 \over \D_g(\T)} {\pa_w \over z-w} \right) R_g^{\a}(\T,w) $$
$$ + { 2 L^{ab} (R_g)_a^{\b} (\T,w) {(\T_b)_{\b}}^{\a}  \over z-w} +
{\rm reg. }  \eqno(A.4) $$
where $\D_{\a}(\T)$ are the $L^{ab}$-broken conformal weights of representation
$\T$. The OPE with $\tT = \tL^{ab} \xx J_a J_b\xx$ is obtained from (A.4) by
the replacement $T \rightarrow \tT, \; \D \rightarrow \tD$ and
$L \rightarrow \tL$.

The OPE (A.4) shows the characteristic affine-Virasoro form (4.3) with the
extra term $\delta R_g^{\a}$ given in (4.5b).

The non-leading term in (A.4), proportional to the composite field
$(R_g)_a^{\a}$, was first seen (for A-S and coset constructions) in Ref.
\cite{h3}. We emphasize here the apparent persistence of the term in the
special case of the A-S construction
$$ T_g(z) R_g^{\a} (\T,w) = \left( { \D_g(\T) \over (z-w)^2} +
{\pa_w \over z-w}  \right) R_g^{\a}(\T,w)
 + { 2 L_g^{ab} (R_g)_a^{\b} (\T,w) {(\T_b)_{\b}}^{\a}  \over z-w} +
{\rm reg. }  \eqno(A.5) $$
where, unless the operator is zero, it contradicts the conventional wisdom that
affine primary fields are Virasoro primary under the A-S construction.

In fact, it is not difficult to check from (3.3b) and (A2.c) that
$$ \langle L_g^{ab} (R_g)_a^{\b}(\T,z) {(\T_b)_{\b}}^{\a} L_g^{cd}
(R_g)_c^{\s}(\bar{\T},w) {(\bar{\T}_d)_{\s}}^{\r} \rangle = 0  \eqno(A.6) $$
so the offending operator is zero for unitary representations of affine
compact $g$. Although it creates only null states, the composite field term
apparently persists for non-unitary constructions, and deserves further study.

\bigskip \bigskip
\centerline{\bf Appendix B: Biprimary fields and the stability condition}
\bigskip
\noindent The original form of the Virasoro biprimary fields \cite{h3}
$$ \phi^{\a} (\bz,z) =z^{L^{(0)}} \bz^{\tilde{L}^{(0)}} \phi_g^{\a}(1)
z^{-L^{(0)}} \bz^{-\tilde{L}^{(0)}} z^{-\D_{\a}} \bz^{-\tD_{\a} } \eqno(B.1)$$
was given only for $h$ and $g/h$, since the Virasoro master equation had not
yet been found. Here, we follow the steps of the orinigal argument to verify
that $\f^{\a}(\bz,z)$ is biprimary for any K-conjugate pair.

We begin with the identities
$$ z\pa_z \f^{\a} (\bz,z) = [L^{(0)}, \f^{\a} (\bz,z)] -
 \D_{\a} \f^{\a} (\bz,z) \eqno(B.2a) $$
$$ z^{-L^{(0)} } L^{(m)} z^{L^{(0)} } = L^{(m)} z^m \eqno(B.2b) $$
$$ [L^{(m)} -L^{(0)} , \f_g^{\a} (1) ] = m \D_{\a} \f_g^{\a} (1) \eqno(B.2c) $$
where (B.2c) is the $L$-stability condition \cite{bh,h3}, which follows from
(4.9a). Then, $\f^{\a}(\bz,z)$ is a Virasoro primary field under $T(z)$ because
$$ \eqalignno{ [& L^{(m)} ,   \f^{\a}(\bz,z)]  = z^{L^{(0)} } \bz^{\tL^{(0)} }
[L^{(m)},\f_g^{\a} (1) ] z^{-L^{(0)} } \bz^{-\tL^{(0)} }
z^{m-\D_{\a} }  \bz^{-\tD_{\a} }  &(B.3a) \cr
& \;\; = z^{L^{(0)} } \bz^{\tL^{(0)} }  \{
[L^{(0)},\f_g^{\a} (1) ] + m \D_{\a} \f_g^{\a} (1) \}
z^{m-L^{(0)}-\D_{\a} } \bz^{-\tL^{(0)}-\tD_{\a} }   &(B.3b)  \cr
& \;\; = z^m (z \pa_z + (m+1)\D_{\a} ) \f^{\a} (\bz,z) &(B.3c) \cr}  $$
where the $L$-stability condition was used in the second step. Similarly,
$\f^{\a} (\bz,z)$ is Virasoro primary under $\tT(z)$ because
$$ [\tL^{(m)} , \f^{\a}(\bz,z) ] =
 \bz^m (\bz \pa_{\bz} + (m+1)\tD_{\a} ) \f^{\a} (\bz,z) \eqno(B.4)  $$
is obtained from (B.1) with the $\tL$-analogues of (B.2).

\bigskip \bigskip
\centerline{\bf Appendix C: Ward identities for $L^{ab}$-broken currents}
\bigskip
\noindent We give the first-order Ward identities  for the
 $L^{ab}$-broken current correlators. Define
$$  \langle \bJ(\bz) \J(z) \rangle_{A_1 \ldots A_n}
\equiv  \langle \J_{A_1} (\bz_1, z_1) \ldots
\J_{A_n} (\bz_n, z_n) \rangle \eqno(C.1a)$$
$$  \langle \bJ(z) \J(z) \rangle_{A_1 \ldots A_n} \equiv
 \langle J_{A_1} ( z_1) \ldots J_{A_n} (z_n) \rangle  \eqno(C.1b)$$
where $\J_A(\bz,z)$ are the biprimary fields (5.2a) of the $L^{ab}$-broken
currents $J_A$. Then the factorized Ward identities
$$\langle \bJ (z) \pa_i \J (z) \rangle_{A_1 \ldots A_n}
= -2L^{CD}\sum_{j \neq i} \left\{
{{f_{CA_i}}^{B_i}  {f_{DA_j}}^{B_j}  \over z_{ij}}
\langle \bJ (z) \J (z) \rangle_{A_1 \,..\, B_i \,..\,B_j \, .. \,A_n} \right.$$
$$ + {G_{CA_i} i{f_{DA_j}}^{B_j}  \over z_{ij}^2}
\langle \bJ (z) \J (z) \rangle_{A_1 \,..\, \hat{A}_i \,..\,B_j \, .. \,A_n}
 - {G_{CA_j} i{f_{DA_i}}^{B_i}  \over z_{ij}^2}
\langle \bJ (z) \J (z) \rangle_{A_1 \,..\, B_i \,..\,\hat{A}_j \, .. \,A_n} $$
$$ \left. + 2 {G_{CA_i} G_{DA_j} \over z_{ij}^3}
\langle\bJ (z) \J (z)\rangle_{A_1 \,..\,\hat{A}_i \,..\,\hat{A}_j \, .. \,A_n}
 \right\} \eqno(C.2)$$
are obtained with (C.1) by choosing $\f_g^{\a} = J_A$ in (6.3). Here, hatted
indices indicate currents which are  not present in the A-S correlators, and
the corresponding right side of $ \langle \pa_i \bJ  \, \J \rangle$ is
obtained from (C.2) by  $L \ra \tL$.

Since ${(T_A^{\rm adjoint})_B}^C = -i {f_{AB}}^C$, the  first term on the
right side of (C.2) is analogous to  the right side of the first-order Ward
identity (8.3a) for the broken affine primaries. The extra inhomogeneous terms,
with hatted indices, arise from the central term of the affine Lie algebra
(2.1b), and such terms are expected generically in the Ward identities of
broken affine secondaries.

These equations are solved identically, before or after factorization, by the
two- and three-point biconformal current correlators in (5.3b) and (5.8).
For the invariant four-point correlators (see eq.(5.11)), we choose the cross
ratios $u$ and $ \bu$ in (9.1b), and the KZ gauge
$$ \g_{12} = \g_{13}  = 0\;\;\;,\;\;\;\; \g_{14} = 2\D_{A_1} \;\;\;,\;\;\;\;
 \g_{23} = \D_{A_1} + \D_{A_2} + \D_{A_3} - \D_{A_4} $$
$$ \g_{24} = -\D_{A_1} + \D_{A_2} - \D_{A_3} + \D_{A_4} \;\;\;,\;\;\;\;
 \g_{34} = -\D_{A_1} - \D_{A_2} + \D_{A_3} + \D_{A_4} \eqno(C.3)$$
$$ \bc_{ij} = \g_{ij} \ve_{\D \ra \tD}  $$
where $\D_A$ is the $L^{ab}$-broken conformal weight of the current $J_A$
and $\tD_A + \D_A = 1$. Then, we obtain the  one-dimensional  non-linear
equations for the invariant current correlators
$$  ( \bY \, \pa Y)_{A_1 A_2A_3 A_4} = -\D_{A_1} (\frac{2}{u^3} G_{A_1A_2}
G_{A_3A_4} + \frac{2}{(u-1)^3} G_{A_1A_3}  G_{A_2A_4}    $$
$$ + \frac{1}{u^2} {f_{A_1 A_2}}^C  f_{A_1 A_3C} + \frac{1}{(u-1)^2}
 {f_{A_1 A_3}}^C  f_{A_4 A_2C} )$$
$$ +\frac{1}{u(u-1)} \sum_C
[\frac{1}{u} (\D_{A_2} -\D_C) {f_{A_1 A_2}}^C  f_{A_1 A_3C}
+\frac{1}{u-1} (\D_{A_3} -\D_C) {f_{A_1 A_3}}^C  f_{A_2 A_4C} $$
$$ - (\D_{A_4} -\D_C) {f_{A_1 A_4}}^C  f_{A_2 A_3C} ] \eqno(C.4)$$
which we have expressed entirely in terms of the $L^{ab}$-broken conformal
weights of the currents. Similarly, the right side of $(\pa \bY \, Y)_A$
is obtained from (C.4) by $\D_A \ra \tD_A$.

\end{document}